\documentclass[english]{article}
\usepackage[T1]{fontenc}
\usepackage[latin1]{inputenc}
\usepackage{geometry}
\usepackage{subfigure}
\usepackage{amsmath}
\usepackage{graphicx}
\usepackage{amssymb}
\usepackage[numbers]{natbib}

\makeatletter
 \newcommand{\lyxaddress}[1]{
   \par {\raggedright #1 
   \vspace{1.4em}
   \noindent\par}
 }

\usepackage{subfigure}
\usepackage{graphicx}
\date{}

\usepackage{babel}
\makeatother
\begin{document}

\title{Pure competition of multiple species during mixed-substrate microbial
growth: Extending the resource-based theory.}

\author{Atul Narang}

\maketitle

\lyxaddress{Department of Chemical Engineering, University of Florida, Gainesville,
FL~32611-6005. narang@che.ufl.edu}

\author{Sergei S. Pilyugin}

\lyxaddress{Department of Mathematics, University of Florida, Gainesville, FL
32611-8105. pilyugin@math.ufl.edu}

\noindent \begin{flushleft}Keywords: Mathematical model, mixed substrate
growth, mixed culture growth, substitutable substrates, coexistence.\end{flushleft}

\begin{abstract}
The simultaneous growth of multiple microbial species is a problem
of fundamental ecological interest. In media containing more than
one growth-limiting substrate, multiple species can coexist. The question
then arises: Can single-species data predict the existence and stability
of mixed-culture steady states in mixed-substrate environments? This
question has been extensively studied with the help of resource-based
models. These studies have shown that the single-species data required
to predict mixed-culture behavior consists of the growth isoclines
and consumption vectors, which in turn are determined from single-substrate
data by making specific assumptions about the kinetics of mixed-substrate
growth. Here, we show that these assumptions are not valid for microbial
growth on mixtures of substitutable substrates. However, the theory
can be developed by determining the growth isoclines and consumption
vectors directly from the mixed-substrate data, thus obviating the
need for specific assumptions about the kinetics of mixed-substrate
growth. We show furthermore that in addition to the growth isoclines
and consumption vectors, the single-species, mixed-substrate data
yields a new family of curves, which we call the \emph{consumption
curve}s. Consideration of the growth isoclines and the consumption
curves yields deeper insights into the behavior of the mixed cultures.
It yields \emph{a priori} bounds on the substrate concentrations achieved
during coexistence, permits the extension of the theory to systems
in which the growth isoclines are non-monotonic, and clarifies earlier
results obtained by considering only the growth isoclines. 
\end{abstract}

\section{Introduction}

In natural water bodies and in many man-made bioreactors, multiple
microbial species thrive in environments containing multiple growth-limiting
substrates. These microbial species interact in diverse and complex
ways, so that the development of a general theory encompassing all
types of interactions is a daunting, if not hopeless, task. We can
begin to make some progress by focusing on systems in which the interaction
between the species is \emph{indirect}, i.e., the specific growth
and nutrient uptake rates of the individual species do not depend
on the densities of the microbial species~\citep{fredrickson81,grover}.
However, the problem remains formidable even if we restrict ourselves
to indirect interactions. For instance, the behavior can be quite
complex, if one or more species excrete metabolic products that stimulate
or inhibit the growth of the other species. If we ignore such indirect
interactions stemming from excretion, we arrive at the problem of
\emph{pure} or \emph{resource competition}. This work is concerned
with the theory of pure competition between multiple microbial species. 

We shall assume that the species in question are growing in a well-stirred
chemostat supplied with a sterile feed having a fixed flow rate and
composition. Natural and man-made water bodies are spatially heterogeneous,
and constantly perturbed by fluctuations in the composition, concentrations,
and flow rates of the nutrients. Yet, we make this idealization because
among all the processes occurring in ecosystems, the biological ones
are the least understood. By confining attention to growth in a chemostat,
we deliberately minimize or eliminate the physicochemical processes
(such as mixing), thus sharpening the focus on the biology of the
system. Furthermore, the restriction to a chemostat ensures that the
theory can be rigorously tested in a laboratory, which is a useful
prelude to field experiments. 

\begin{figure}
\begin{center}\subfigure[]{\includegraphics[%
  width=7cm,
  height=5cm]{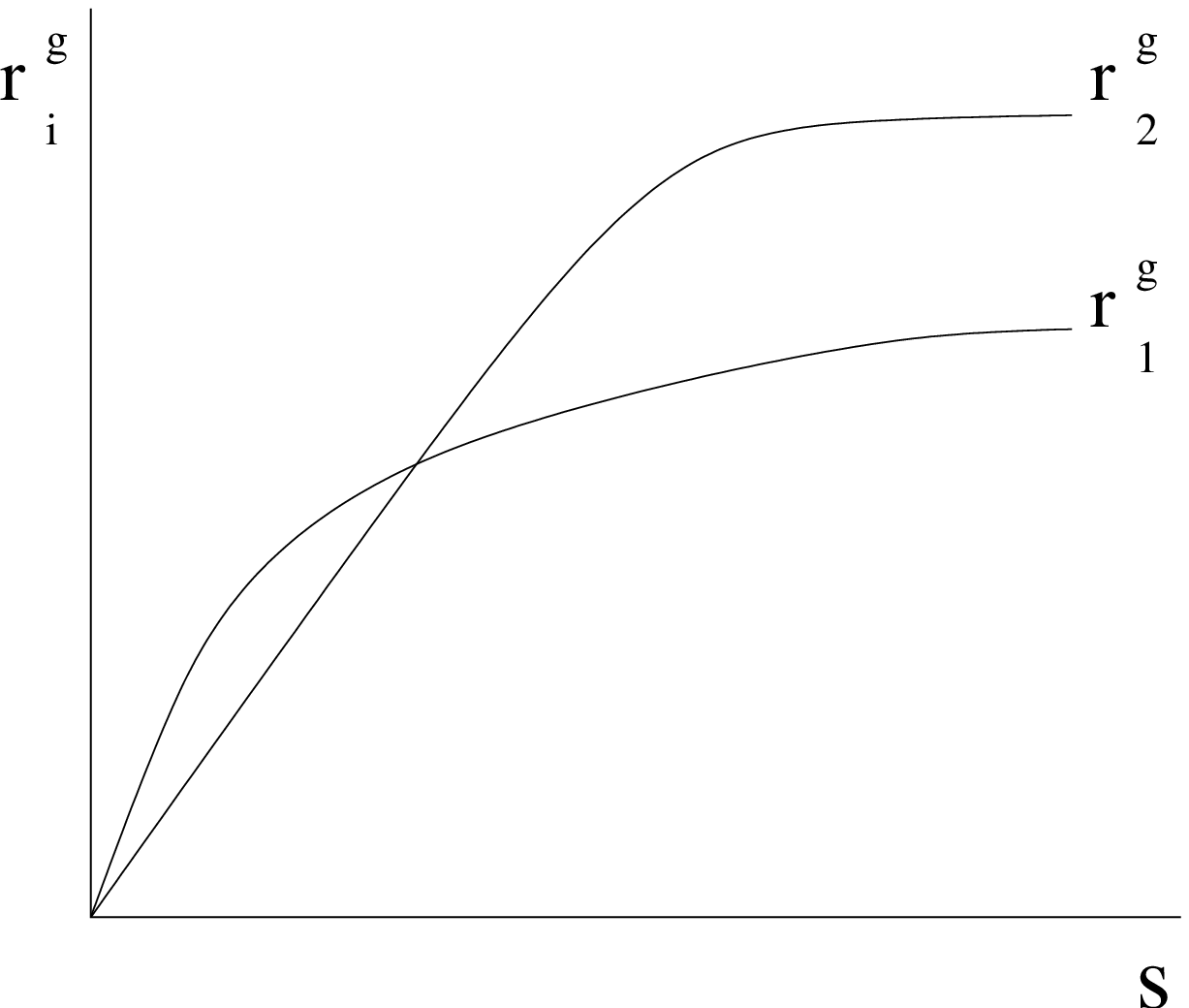}}\hspace{0.1in}\subfigure[]{\includegraphics[%
  width=7cm,
  height=5cm]{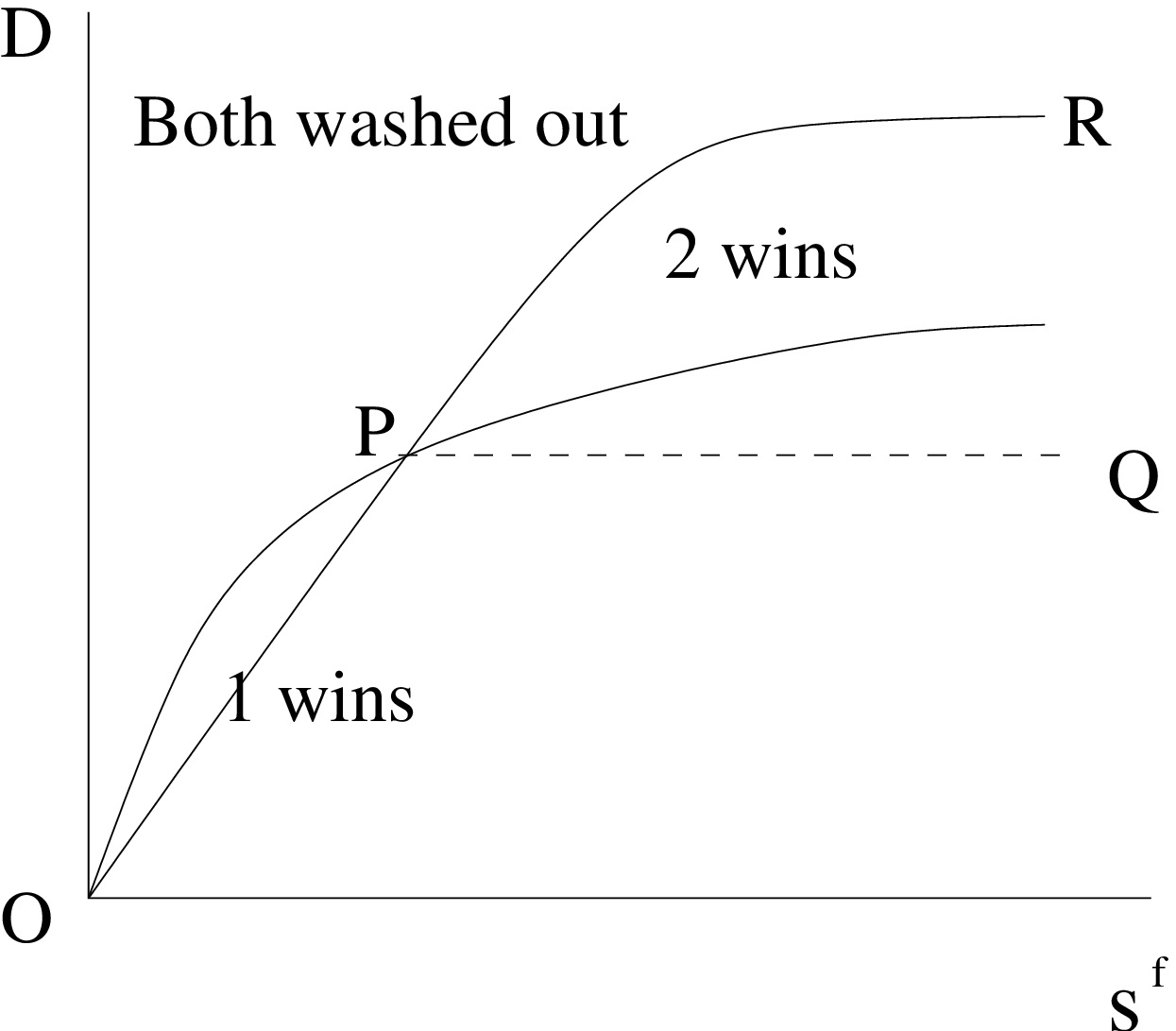}}\end{center}

\caption{\label{f:SingleSubstrateCase}In environments containing a single
growth-limiting substrate, the steady states of mixed-culture growth
are completely determined by the single-species data. (a)~Variation
of the specific growth rates of the two species, denoted $r_{1}^{g}$
and $r_{2}^{g}$, as a function of the substrate concentration, $s$.
(b)~The corresponding mixed-culture steady states at any dilution
rate, $D$, and feed concentration, $s^{f}$. Only species~1 survives
if $D$ and $s^{f}$ lie the region OPQ, and only species~2 survives
if $D$ and $s^{f}$ are in the region QPR. Both species wash out
for all other $D$ and $s^{f}$.}
\end{figure}

Intuition suggests that when multiple species engage in pure competion,
it should be possible to predict their behavior \emph{a priori} from
single-species experiments. For, under these conditions, each species
sees only the exogenous substrates in its environment --- the other
species are perceived through the effects they exert on the environment
by consumption of the substrates. But the substrate consumption patterns
of each species can be determined by performing single-species experiments.
Thus, it seems plausible that given appropriate single-species data,
one should be able to predict the behavior of mixed (multi-species)
cultures.

The foregoing intuition is borne out when multiple species compete
for a \emph{single} growth-limiting substrate~\citep{aris77,hsu78,hsu77,powell58,wolkowicz92}.
Under these conditions, the single-species data required to predict
their behavior in mixed cultures consists of the \emph{growth curves},
which define the variation of the specific growth rate, $r_{i}^{g}$,
as a function of the growth-limiting substrate concentration, $s$
(Figure~\ref{f:SingleSubstrateCase}a). Given these curves, we can
predict the mixed-culture steady states at any dilution rate, $D$,
and feed concentration, $s^{f}$. It suffices to generate Figure~\ref{f:SingleSubstrateCase}b
from Figure~\ref{f:SingleSubstrateCase}a by replacing the $s$-
and $r_{i}^{g}$-axes in Figure~\ref{f:SingleSubstrateCase}a with
$s^{f}$ and $D$, respectively, and drawing a horizontal line passing
through the intersection point, if any, of the two growth curves.
Figure~\ref{f:SingleSubstrateCase}b predicts that if the feed concentration
is held fixed at a sufficiently high value, and the dilution rate
is increased, only species~1 survives at low dilution rates, only
species~2 survives at intermediate dilution rates, and both species
wash out at high dilution rates~\citep{fredrickson77a}. This prediction
has been confirmed by numerous experiments (reviewed in~\citep[Chapter 3]{grover}
and~\citep{veldkamp77}).

\begin{figure}
\begin{center}\includegraphics[%
  width=7cm,
  height=5cm]{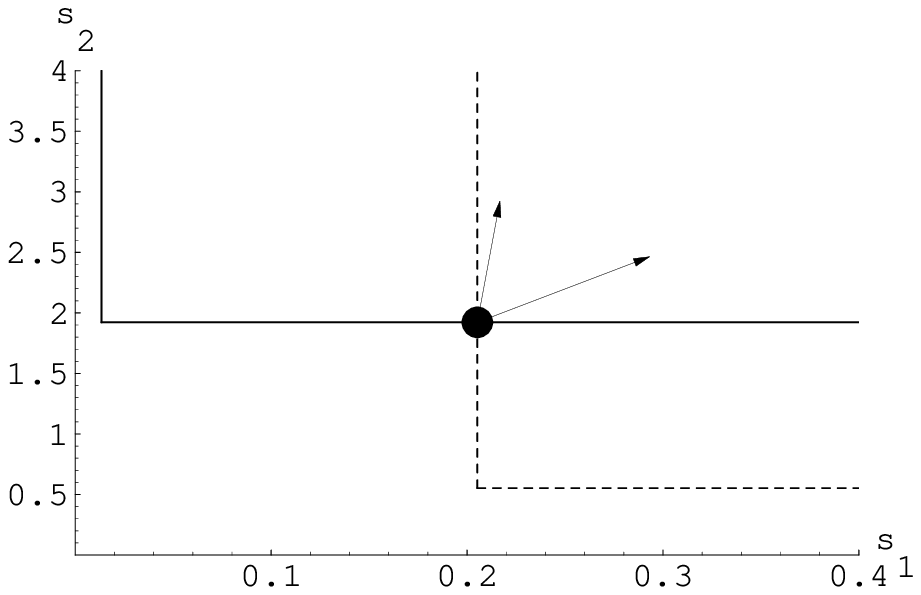}\hspace{0.1in}\includegraphics[%
  width=7cm,
  height=5cm]{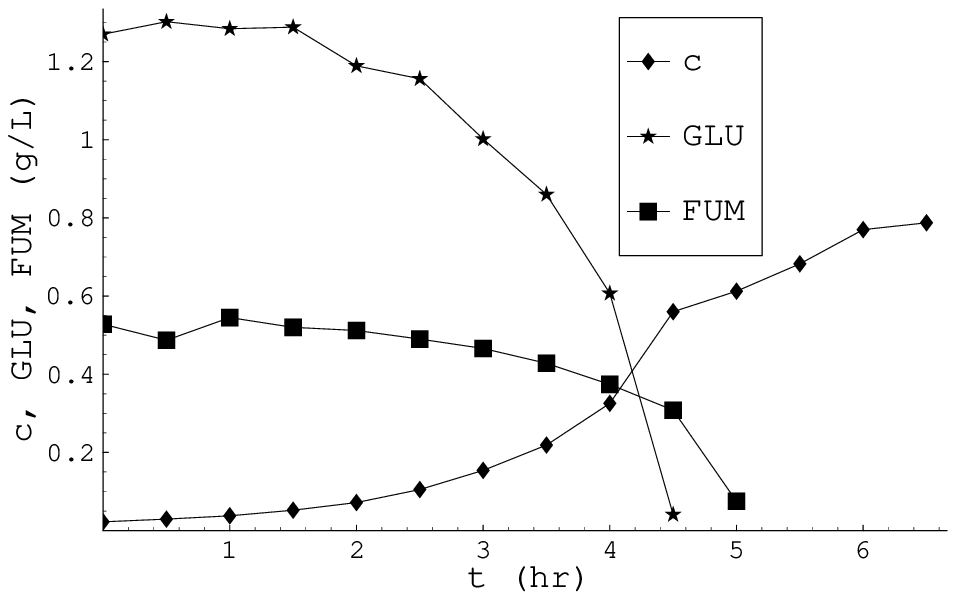}\end{center}

\caption{\label{f:TilmanCriterion}(a) Growth of \emph{A.~formosa} (species~1)
and \emph{C.~meneghianiana} (species~2) on a mixture of phosphate
($s_{1}$) and silicate ($s_{2}$) ~\citep{tilman80,tilman}. The
full and dashed lines show the \emph{growth isoclines} for species~1
and~2 at $D=0.25$~1/day. They intersect at substrate concentrations
representing a potential coexistence steady state. The arrows show
the corresponding \emph{consumption vectors,} $(r_{11}^{s},r_{12}^{s})$
and $(r_{21}^{s},r_{22}^{s})$, for the two species, where $r_{ij}^{s}$
is the specific uptake rate of the $j^{{\rm th}}$ substrate by the
$i^{{\rm th}}$ species. The growth isoclines and consumption vectors
were calculated by assuming that $r_{i}^{g}$ follows Liebig's law,
and $r_{ij}^{s}=r_{i}^{g}/Y_{ij}$, where $Y_{ij}$ denotes the single-substrate
specific growth rate and yield of the $i^{{\rm th}}$ species on $S_{j}$.
(b) Diauxic growth of \emph{E. coli} K12 on a mixture of glucose and
fumarate (from~\citep{narang97a}). Consumption of fumarate does
not begin until almost complete exhaustion of glucose.}
\end{figure}

It is of considerable interest to determine if single-species data
is also sufficient for predicting the behavior of mixed cultures limited
by \emph{multiple} growth-limiting substrates. A convenient starting
point is the problem of mixed-culture growth in the presence of two
growth-limiting substrates. This question has been addressed by a
considerable body of work (reviewed in~\citep[Chapter 2]{grover}).
The crux of these theoretical developments is that in two-substrate
environments, no more than two species can coexist at steady state.
Moreover, the existence and stability of the coexistence steady state
at any given dilution rate and feed concentrations can be predicted
\emph{a priori} if we know two pieces of information derived from
single-species data, namely~\citep{tilman80,tilman}

\begin{enumerate}
\item The \emph{growth isocline} for each species, which is the locus of
all substrate concentrations at which the species can maintain a specific
growth rate equal to the given dilution rate. These curves determine
the feasibility of coexistence steady states during mixed-culture
growth --- such steady states exist only if the growth isoclines for
the two species intersect.
\item The two \emph{consumption vectors} at the coexistence steady state(s),
which represent the substrate uptake rates of the two species. These
vectors determine whether the coexistence steady states are experimentally
observable. More precisely, a coexistence steady state is stable only
if the feed concentrations lie in the cone generated by the consumption
vectors.
\end{enumerate}
In the literature, the growth isoclines and consumption vectors are
determined by making an additional hypothesis. Specifically, it is
assumed that the growth and substrate uptake rates during mixed-substrate
growth are completely determined by the growth and substrate uptake
rates during \emph{single-substrate} growth. This assumption works
quite well if the two growth-limiting substrates are \emph{essential}
or \emph{heterologous} (satisfy distinct nutritional requirements).
In this case, the growth of both species is limited by a single substrate
at all but a thin {}``band'' of substrate concentrations. This thin
band is often idealized as a curve by assuming that growth follows
Liebig's law, $r_{i}^{g}=\min\{ r_{i1}^{g}(s_{1}),r_{i2}^{g}$($s_{2})\}$,
where $r_{ij}^{g}(s_{j})$ denotes the single-substrate specific growth
rate of the $i^{{\rm th}}$ species on $S_{j}$. Consequently, the
growth isoclines and consumption vectors can be determined by the
single-substrate data (Figure~\ref{f:TilmanCriterion}a). Many predictions
of the theory for essential substrates have been verified (see~\citep{Miller2005}
for a recent review).

The very same assumption --- single-substrate data determines the
mixed-substrate behavior --- is also made when the growth-limiting
substrates are \emph{substitutable} or \emph{homologous}, i.e., satisfy
identical nutritional requirements (Figure~\ref{f:TilmanCriterion}b).
More precisely, it is assumed that the specific growth rate of the
$i^{{\rm th}}$ species, $r_{i}^{g}$, has the form\begin{equation}
r_{i}^{g}(s_{1},s_{2})=r_{i1}^{g}(s_{1})+r_{i2}^{g}(s_{2})\label{eq:AdditiveGrowthRates}\end{equation}
where $r_{i1}^{g}(s_{1})$ and $r_{i2}^{g}(s_{2})$ represent the
\emph{single-substrate} specific growth rates of the $i^{{\rm th}}$
species on $S_{1}$ and $S_{2}$, respectively, and are often assumed
to obey Monod kinetics~\citep{Rothhaupt1988}. But this assumption
is not consistent with the experimental data for microbial growth
on mixtures of substitutable substrates. Indeed

\begin{enumerate}
\item During batch growth on such mixtures, when the exogenous concentrations
of both substrates are relatively high, one frequently observes \emph{diauxic}
growth~\citep{egli95,kovarova98}. During such growth, the microbes
preferentially consume only one of the substrates, and it is only
after exhaustion of this {}``preferred'' substrate that they start
consuming the other {}``less preferred'' substrate (Figure~\ref{f:TilmanCriterion}b).
Thus, the specific growth rate observed in such mixed-substrate experiments
is identical to the maximum specific growth on the {}``preferred''
substrate. In contrast, assumption~(\ref{eq:AdditiveGrowthRates})
implies that the mixed-substrate specific growth rate equals the sum
of the maximum specific growth rates on each of the substrates. 
\item During mixed-substrate growth in continuous cultures, the cells stop
consuming the {}``less preferred'' substrate at a sufficiently high
dilution rate (in~Figure~\ref{f:ConsumptionCurves1}a, for example,
there is no consumption of methanol for $D>0.35$~1/hr). The model
predicts that this dilution rate is equal to the maximum specific
growth rate on the {}``less preferred'' substrate. But experiments
show that this dilution rate is always larger than the maximum specific
growth on the {}``less preferred'' substrate~\citep{egli95}.%
\footnote{In Tilman's terminology, the substrates are \emph{perfectly substitutable}
resources at low dilution rates, and \emph{switching} resources at
high dilution rates. However, the dilution rate at which the substrates
undergo this transition is always greater than the washout dilution
rate for the {}``less preferred'' substrate.%
}
\end{enumerate}
These discrepancies between the model and experiments arise because
assumption~(\ref{eq:AdditiveGrowthRates}) ignores the fact that
the growth rate supported by a substrate is inhibited by the other
substrate. Indeed, Monod and coworkers showed that diauxic growth
occurs because one of the substrates completely blocks the uptake,
and hence the growth, on the other substrate. Thus, a more accurate
representation of the specific growth must have the form\[
r_{i}^{g}(s_{1},s_{2})=r_{i1}^{g}(s_{1},s_{2})+r_{i2}^{g}(s_{1},s_{2}),\]
where $r_{i1}^{g}(s_{1},s_{2})$ and $r_{i2}^{g}(s_{1},s_{2})$ are
decreasing functions of $s_{2}$ and $s_{1}$, respectively. Since
these inhibitory effects are manifested only in the presence of \emph{both}
substrates, it is impossible to infer mixed-substrate behavior from
single-substrate experiments, except at low growth rates or dilution
rates when the inhibitory effects are negligibly small. Hence, any
theory of mixed-culture growth on substitutable substrates must be
developed without invoking assumption~(\ref{eq:AdditiveGrowthRates}). 

In this work, we make no attempt to infer mixed-substrate kinetics
from single-substrate data. Instead, we show that the growth isoclines
and the consumption vectors can be calculated directly from single-species,
mixed-substrate experiments. Thus, we develop a complete theory that
appeals only to the data --- it makes no assumptions about the relationship
between single- and mixed-substrate behavior. We show furthermore
that in addition to the growth isoclines, it is useful to consider
the \emph{consumption curves}, i.e., the locus of substrate concentrations
at which the rates of substrate supply and consumption are perfectly
balanced. Consideration of the consumption curves along with the growth
isoclines yields deeper insights into the steady states of mixed cultures.
Specifically, we obtain

\begin{enumerate}
\item \emph{A priori} bounds on the substrate concentrations that can be
attained in mixed cultures.
\item Information about the stability of the steady states even if the growth
isoclines are non-monotonic.
\item Additional insight into the nature of the transitions (bifurcations)
that occur when the feed concentrations are changed at fixed dilution
rate (or the dilution rate is changed at fixed feed concentrations).
\end{enumerate}
Fortunately, the determination of consumption curves requires no additional
effort. Given the growth isoclines and consumption vectors calculated
from single-species experiments, the consumption curves can be determined
without further experiments. Thus, the benefits of consumption curves
can be realized at no further cost in terms of experiments.

The theory developed in this work rests upon three assumptions ---
pure competition, mutual inhibition of substrate uptake rates, and
constant yields. There is substantial experimental evidence supporting
the last two assumptions. These have been comprehensively reviewed
by Egli and coworkers~\citep{egli95,kovarova98}, and we refer the
reader to the references cited in these reviews for experimental tests
of these assumptions. Here, we focus on specific experiments aimed
at testing the validity of the first hypothesis, namely, pure competition.
Most microbes excrete metabolic products, particularly at high dilution
rates~\citep{holms86_1}. It is important to ensure that the system
in question displays pure competition despite the release of excretory
products.

Thus, the theory presented here achieves three goals 

\begin{enumerate}
\item It shows that the specific assumptions made in order to infer mixed-substrate
behavior from single-substrate data are unnecessary. The single-species,
mixed-substrate data is sufficient for predicting all features of
mixed-culture growth. 
\item It extends the resource-based approach by revealing the additional
information embedded in the consumption curves. 
\item It suggests experiments aimed at testing the hypotheses of the theory.
\end{enumerate}
The paper is organized as follows. In Section~\ref{s:Results}, we
describe the model, define the growth isoclines and the consumption
curves, and show that they determine the properties of all the mixed-culture
steady states. In Section~\ref{s:Discussion}, we describe experiments
that may be used to test the model hypotheses and predictions. Finally,
we summarize the conclusions in Section~\ref{s:Conclusions}.

\section{\label{s:Results}Results}

\subsection{The model}

In a chemostat fed with sterile nutrients, the mass balances for the
two growth-limiting substrates and species yield the equations~\citep{ballyk93,leon75,tilman80}\begin{align}
\frac{ds_{j}}{dt} & =D(s_{j}^{f}-s_{j})-c_{1}r_{1j}^{s}-c_{2}r_{2j}^{s},\quad j=1,2,\label{eq:MCunstructuredS}\\
\frac{dc_{i}}{dt} & =\left(r_{i}^{g}-D\right)c_{i},\quad i=1,2,\label{eq:MCunstructuredC}\end{align}
 where $c_{i}$ (gdw/L) and $s_{j}$~(g/L) denote the concentrations
in the chemostat of the $i^{\mathrm{th}}$~species and the $j^{\mathrm{th}}$~substrate;
$D$~(1/hr) denotes the dilution rate; $s_{j}^{f}$~(g/L) denotes
the feed concentration of the $j^{\mathrm{th}}$~substrate; $r_{i}^{g}$~(1/hr)
denotes the specific growth rate of the $i^{\mathrm{\mathrm{th}}}$
species; and $r_{ij}^{s}$~(g/gdw-hr) denotes the specific uptake
rate of the $j^{\mathrm{th}}$~substrate by the $i^{\mathrm{\mathrm{th}}}$~species.
To complete the model, we must specify the kinetics of growth ($r_{i}^{g}$)
and substrate uptake ($r_{ij}^{s}$). 

We assume that the specific growth and substrate uptake rates are
functions of the exogenous substrate concentrations only, i.e.,\begin{equation}
r_{i}^{g}(s_{1},s_{2}),\; r_{ij}^{s}(s_{1},s_{2}).\label{eq:HypothesisUniqueness}\end{equation}
 This is the mathematical representation of the assumption that the
competition is pure, i.e., the growth and substrate uptake rates are
independent of the cell densities. 

Insofar as the specific uptake rates are concerned, we assume that
there is no substrate uptake in the absence of the substrate, i.e.,\begin{equation}
r_{i1}^{s}(0,s_{2})=r_{i2}^{s}(s_{1},0)=0,\label{eq:HypothesisZeroUptake}\end{equation}
 and each substrate stimulates its own uptake\begin{equation}
\frac{\partial r_{i1}^{s}}{\partial s_{1}}(s_{1},s_{2})>0,\frac{\partial r_{i2}^{s}}{\partial s_{2}}(s_{1},s_{2})>0.\label{eq:HypothesisMChomo}\end{equation}
 Thus, we preclude the self-inhibitory effects often observed at relatively
high substrate concentrations. We assume furthermore that each substrate
inhibits the uptake of the other substrate, i.e.,\begin{equation}
\frac{\partial r_{i1}^{s}}{\partial s_{2}}(s_{1},s_{2})<0,\frac{\partial r_{i2}^{s}}{\partial s_{1}}(s_{1},s_{2})<0.\label{eq:HypothesisMChetero}\end{equation}
 This \emph{mutual inhibition} is characteristic of growth on mixtures
of substitutable substrates. It should be emphasized this assumption
does not imply that each substrate directly inhibits the uptake of
the other substrate. In reality, each substrate inhibits the synthesis
of the transport enzymes for the other substrate~\citep{stulke99}.
However, since these enzymes are not variables of the model, (\ref{eq:HypothesisMChetero})
captures the net effect of the mutual inhibition.

It remains to specify the properties of the specific growth rates.
To this end, let $Y_{ij}$ denote the yield of the $i^{\mathrm{th}}$
species on $S_{j}$. Since the substrates are substitutable, the specific
growth rate has the form\begin{equation}
r_{i}^{g}(s_{1},s_{2})=Y_{i1}r_{i1}^{s}(s_{1},s_{2})+Y_{i2}r_{i2}^{s}(s_{1},s_{2})\label{eq:HypothesisAdditiveYield}\end{equation}
 We assume that the yields are constant.

The relations (\ref{eq:HypothesisUniqueness}--\ref{eq:HypothesisAdditiveYield})
embody all the assumptions required for the development of our theory.
However, in previous work, it has often been assumed that $\partial r_{i}^{g}/\partial s_{1},\partial r_{i}^{g}/\partial s_{2}>0$.
Thus, it seems to appropriate to understand the conditions under which
this additional assumption is manifested by our model. Now, (\ref{eq:HypothesisAdditiveYield})
implies that \[
\frac{\partial r_{i}^{g}}{\partial s_{1}}=Y_{i1}\frac{\partial r_{i1}^{s}}{\partial s_{1}}+Y_{i2}\frac{\partial r_{i2}^{s}}{\partial s_{1}},\;\frac{\partial r_{i}^{g}}{\partial s_{2}}=Y_{i1}\frac{\partial r_{i1}^{s}}{\partial s_{2}}+Y_{i2}\frac{\partial r_{i2}^{s}}{\partial s_{2}}.\]
In general, nothing can be said about the signs of $\partial r_{i}^{g}/\partial s_{1}$
and $\partial r_{i}^{g}/\partial s_{2}$ because even though each
substrate stimulates growth by promoting its own uptake, it also inhibits
growth by depressing the uptake of the other substrate. For instance,
the sign of $\partial r_{i}^{g}/\partial s_{1}$ is indeterminate
because $\partial r_{i1}^{s}/\partial s_{1}>0$ ($S_{1}$ promotes
its own uptake), but $\partial r_{i2}^{s}/\partial s_{1}<0$ ($S_{1}$
inhibits the uptake of $S_{2}$). It is only when the intensity of
mutual inhibition, represented by the magnitudes of $\partial r_{i1}^{s}/\partial s_{2}$
and $\partial r_{i2}^{s}/\partial s_{1}$, is sufficiently small that
both substrates stimulate the specific growth rate, i.e.,\begin{equation}
\frac{\partial r_{i}^{g}}{\partial s_{1}}(s_{1},s_{2})>0,\frac{\partial r_{i}^{g}}{\partial s_{2}}(s_{1},s_{2})>0.\label{eq:HypothesisWeakInhibition}\end{equation}
 Accordingly, we shall refer to this as the case of \emph{weak} mutual
inhibition. While this is true in many instances, we show below that
there are examples of mixed-substrate growth in which the specific
growth rate is a decreasing function of one of the substrate concentrations.
In what follows, we begin by assuming weak mutual inhibition. Later
on, however, we shall relax this assumption.

\subsection{There are three types of steady states }

It follows from equations~(\ref{eq:MCunstructuredS}--\ref{eq:MCunstructuredC})
that the model has 3 types of steady states

\begin{enumerate}
\item The \emph{trivial} steady state at which neither species survives
in the chemostat ($c_{1}=c_{2}=0$). We denote it by $\phi_{00}$. 
\item The \emph{semitrivial} steady state at which only one of the species
survives in the chemostat. There are two types of semitrivial steady
states, $c_{1}>0,c_{2}=0$ and $c_{1}=0,c_{2}>0$. We denote them
by $\phi_{10}$ and $\phi_{01}$, respectively. 
\item The \emph{nontrivial} steady state at which both species coexist,
i.e., $c_{1}>0,c_{2}>0$. We denote such steady states by $\phi_{11}$. 
\end{enumerate}
Our goal is to show that given the validity of the model described
above, the steady state data obtained from single-species experiments
is sufficient for predicting the properties of the mixed-culture steady
states.

\subsection{Definition of the growth isoclines and consumption curves}

We begin by defining the single-species data required to infer the
properties of the mixed-culture steady states. To this end, we show
first of all that the determination of the growth isoclines and consumption
vectors from single-species experiments is, in effect, a method for
characterizing the growth and substrate consumption kinetics of a
given species. We then show that the determination of consumption
curves and consumption vectors from single-species experiments is
another way of obtaining the same information. Finally, we establish
the equivalence of the two methods by showing that given the growth
isoclines and consumption vectors, we can infer the consumption curves;
conversely, the growth isoclines can be inferred from the consumption
curves. Thus, either one of the two methods can be used to obtain
the growth isoclines, consumption vectors, and consumption curves.

\begin{figure}[t]
\begin{center}\subfigure[]{\includegraphics[%
  width=7cm,
  height=5cm]{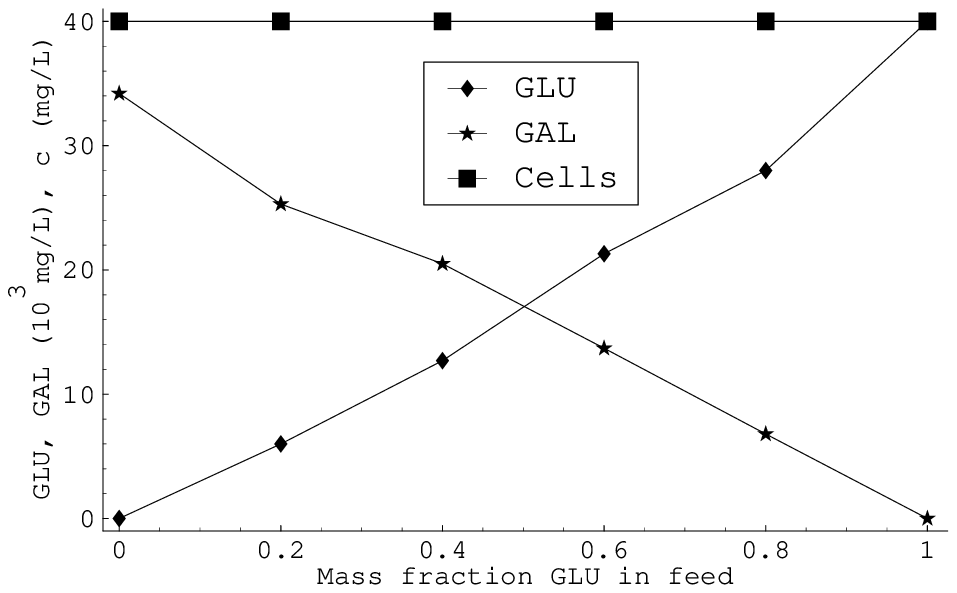}}\hspace{0.5in}\subfigure[]{\includegraphics[%
  width=7cm,
  height=5cm]{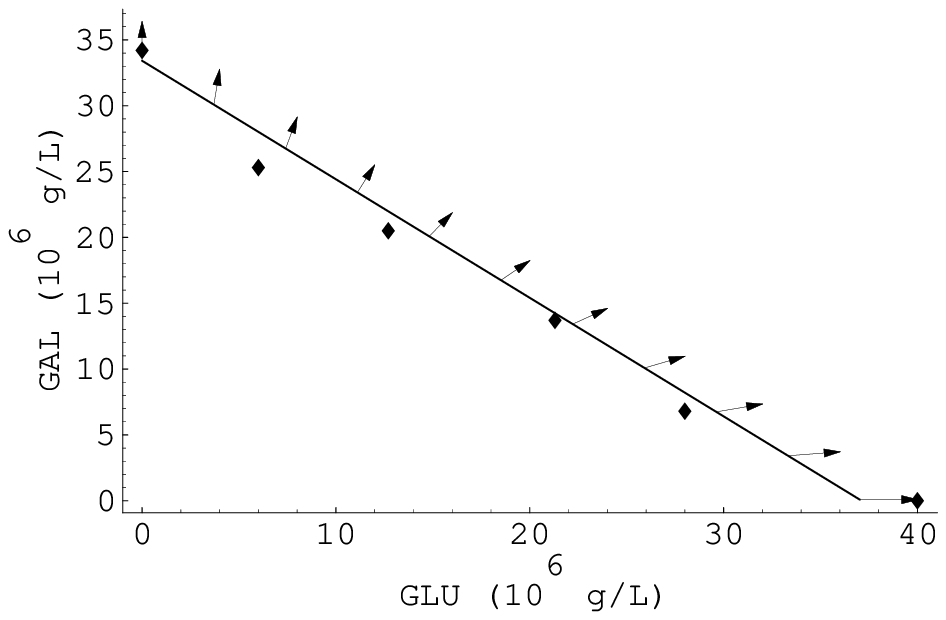}}\end{center}

\caption{\label{f:GrowthIsoclines1}Construction of the growth isoclines:
(a)~The steady state substrate concentrations and cell densities
during growth of \emph{E. coli} ML308 on a mixture of glucose and
galactose at $D=0.3$~1/hr and varying feed concentrations (from~\citep{lendenmann96}).
The feed concentrations were varied by fixing the total sugar concentration
at 100~mg/L and altering the composition. (b) The corresponding growth
isocline and the associated specific substrate consumption vectors. }
\end{figure}

\paragraph{The growth isoclines and consumption vectors capture the growth and
substrate consumption kinetics}

The growth isocline of the $i^{\mathrm{\mathrm{th}}}$ species at
dilution rate, $D$, denoted $\Upsilon_{i}(D)$, is the locus of all
steady state substrate concentrations obtained when this species alone
is grown on a mixture of $S_{1}$ and $S_{2}$ at the \emph{fixed
dilution rate,} $D$, and varying feed concentrations. 

The upper panel of Figure~\ref{f:GrowthIsoclines1} illustrates the
construction of the growth isocline from single-species data. Figure~\ref{f:GrowthIsoclines1}a
shows the steady state sugar concentrations obtained during growth
of of \emph{E. coli} ML308 on a mixture of glucose and galactose when
the feed composition is changed at fixed dilution rate ($D=0.3$~1/hr)
and total feed concentration ($s_{1}^{f}+s_{2}^{f}=100$~mg/L). The
growth isocline at $D=0.3$~1/hr is obtained from this data by plotting
the pairs of glucose and galactose concentrations at various feed
compositions on the plane of substrate concentrations (Figure~\ref{f:GrowthIsoclines1}b).
Evidently, the same procedure can be performed at various dilution
rates, thus generating a family of growth isoclines parametrized by
the dilution rate. 

Having determined the family of growth isoclines, one can draw vectors
perpendicular to the growth isoclines, and pointing in the direction
of increasing $D$. The magnitude of these vectors is arbitrary, but
their direction is well-defined. For reasons given below, we shall
refer to these vectors as the \emph{growth limitation vectors}.

To understand the biological meaning of the experimentally determined
growth isoclines and growth limitation vectors, it is useful to derive
their mathematical counterparts in terms of the model. To this end,
observe that the steady states for single-species growth of the $i^{{\rm th}}$
species satisfy the equations\begin{eqnarray}
D(s_{1}^{f}-s_{1}) & = & c_{i}r_{i1}^{s}(s_{1},s_{2}),\label{eq:ConsumptionCurve1}\\
D(s_{2}^{f}-s_{2}) & = & c_{i}r_{i2}^{s}(s_{1},s_{2}),\label{eq:ConsumptionCurve2}\\
r_{i}^{g}(s_{1},s_{2}) & = & D.\label{eq:MCgrowthIsocline}\end{eqnarray}
Evidently, equation~(\ref{eq:MCgrowthIsocline}) defines the growth
isocline, since it determines the locus of all steady state concentrations
attained at fixed $D$. Thus, the growth isocline for the $i^{{\rm th}}$
species at dilution rate, $D$, is the locus of all substrate concentrations
at which the specific growth rate of the $i^{{\rm th}}$ species equals
the dilution rate. 

Since the growth isocline is given by~(\ref{eq:MCgrowthIsocline}),
it is clear that the growth limitation vector defined above is parallel
to the gradient of the specific growth rate\begin{equation}
\triangledown r_{i}^{g}(s_{1},s_{2})\equiv\left(\frac{\partial r_{i}^{g}}{\partial s_{1}},\frac{\partial r_{i}^{g}}{\partial s_{2}}\right).\label{eq:GrowthLimitationVector}\end{equation}
It follows that the slope of the growth limitation vector provides
a measure of the extent to which growth is limited by $S_{1}$ or
$S_{2}$. Indeed, if the slope of this vector is zero, then so is
the slope of $\nabla r_{i}^{g}(\mathbf{s})$. In this case, (\ref{eq:GrowthLimitationVector})
implies that $\partial r_{i}^{g}/\partial s_{2}=0$, i.e., growth
is limited exclusively by $S_{1}$. More generally, if the slope of
the growth limitation vector is small but not zero, we may say that
growth is more limited by $S_{1}$ rather than $S_{2}$. Similarly,
if the slope of the growth limitation vector is large, growth is more
limited by $S_{2}$ rather than $S_{1}$. 

\begin{figure}[t]
\begin{center}\subfigure[]{\includegraphics[%
  width=7cm,
  height=5cm]{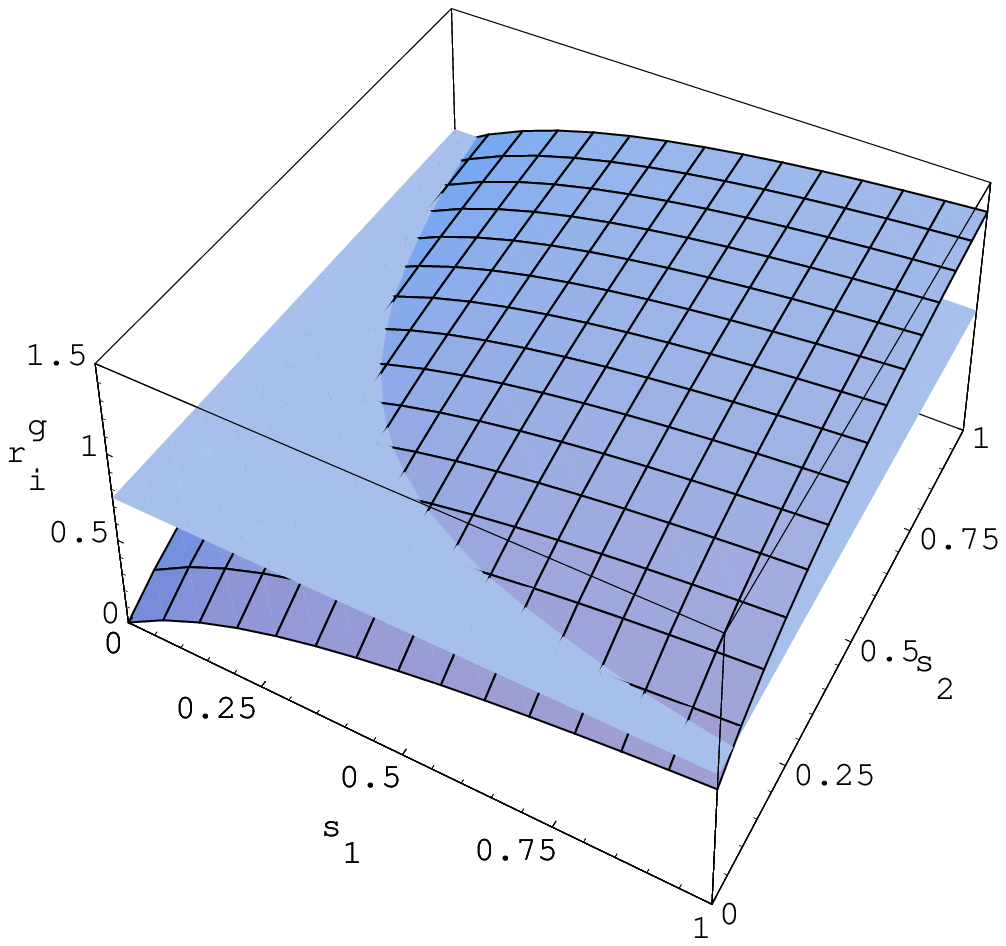}}\hspace{0.5in}\subfigure[]{\includegraphics[%
  width=7cm,
  height=5cm]{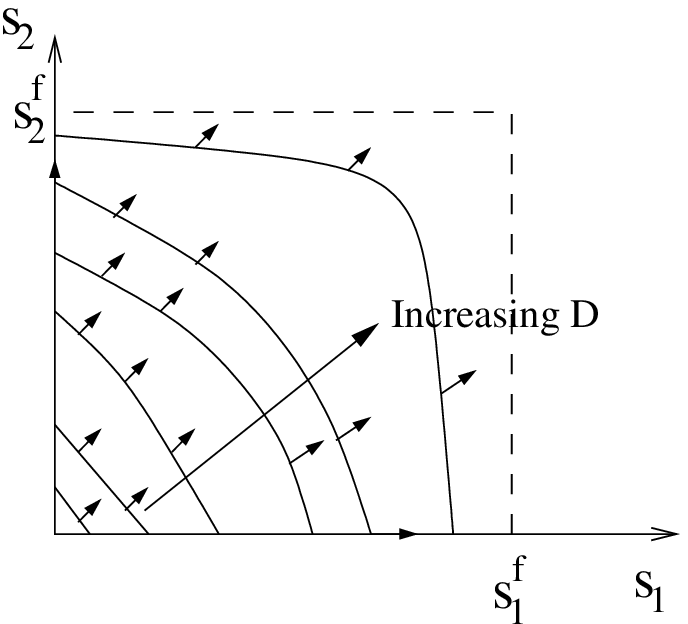}}\end{center}

\caption{\label{f:GrowthIsoclines2}Geometry of the growth isoclines: (a)~The
figure shows the typical surface of $r_{i}^{g}(s_{1},s_{2})$ in the
case of weak mutual inhibition, when $r_{i}^{g}(s_{1},s_{2})$ is
an increasing function of $s_{1}$ and $s_{2}$. Also shown in the
figure is a plane of height, $0.75$~1/hr, which intersects the surface
of $r_{i}^{g}(s_{1},s_{2})$. The growth isocline of the $i^{{\rm th}}$
species at $D=0.75$~1/hr is the projection of this curve of intersection
onto the $s_{1}s_{2}$-plane. It is clear from the figure that this
growth isocline is a decreasing curve on the $s_{1}s_{2}$-plane.
At substrate concentrations above the growth isocline, $r_{i}^{g}(s_{1},s_{2})>0.75$~1/hr;
at substrate concentrations below the growth isocline, $r_{i}^{g}(s_{1},s_{2})<0.75$~1/hr.
Moreover, the higher the value of $D$, the farther the corresponding
growth isocline from the origin of the $s_{1}s_{2}$-plane. (b)~Consequently,
the growth isoclines at various $D$ form a family of decreasing curves
(shown as full lines). The dashed lines show a hypothetical path along
which the feed concentrations are changed in order to generate the
growth isoclines.}
\end{figure}

The shape of the growth isoclines reflects the geometry of the surface
of $r_{i}^{g}(s_{1},s_{2})$. If the mutual inhibition is weak, the
specific growth rate is an increasing function of both substrate concentrations,
so that the growth isocline at any $D$ is a decreasing curve (Figure~\ref{f:GrowthIsoclines2}).
At substrate concentrations below the growth isocline, the specific
growth rate is less than the dilution rate ($r_{i}^{g}<D$); at substrate
concentrations above the growth isocline, it exceeds the dilution
rate ($r_{i}^{g}>D$). Moreover, the higher the value of $D$, the
further the distance of the growth isocline from the origin.

Thus, the construction of growth isoclines from single-species experiments
reveals the behavior of the function, $r_{i}^{g}(s_{1},s_{2})$. But
the single-species experiments also yield the specific substrate uptake
rates. Indeed, (\ref{eq:ConsumptionCurve1}--\ref{eq:ConsumptionCurve2})
imply that \begin{equation}
r_{i1}^{s}(s_{1},s_{2})=\frac{D(s_{1}^{f}-s_{1})}{c_{i}},\; r_{i2}^{s}(s_{1},s_{2})=\frac{D(s_{2}^{f}-s_{2})}{c_{i}}.\label{eq:SpecificUptakeRates}\end{equation}
Since the operating conditions ($D$, $s_{1}^{f}$, $s_{2}^{f}$)
are known, and the steady state concentrations ($s_{1}$, $s_{2}$,
$c_{i}$) are measured, we can calculate $r_{i1}^{s}(s_{1},s_{2})$
and $r_{i2}^{s}(s_{1},s_{2})$ at every pair of substrate concentrations
used to construct the growth isocline. It is convenient to represent
the two specific uptake rates by a vector, $\mathbf{r_{i}^{s}}(\mathbf{s})\equiv\left(r_{i1}^{s}(\mathbf{s}),r_{i2}^{s}(\mathbf{s})\right)$,
based at the corresponding steady state substrate concentrations (see
Figure~\ref{f:GrowthIsoclines1}b). We shall refer to $\mathbf{r_{i}^{s}}(\mathbf{s})$
as the \emph{specific substrate consumption vector} of the $i^{{\rm th}}$
species. 

The slope of $\mathbf{r_{i}^{s}}(\mathbf{s})$ reveals the substrate
consumption behavior of the $i^{{\rm th}}$ species. If the slope
of $\mathbf{r_{i}^{s}}(\mathbf{s})$ is small, then $r_{i2}^{s}(\mathbf{s})\ll r_{i1}^{s}(\mathbf{s})$,
which means that the $i^{{\rm th}}$ species consumes more $S_{1}$
rather than $S_{2}$. If the slope is large, it consumes more $S_{2}$
rather than $S_{1}$.

The generation of Figure~\ref{f:GrowthIsoclines2}b from single species
data provides complete information about the kinetics of growth and
substrate consumption. Given any $(s_{1},s_{2})$ on a growth isocline,
the corresponding specific growth rate is equal to the dilution rate
at which the growth isocline was constructed, and the specific substrate
uptake rates are given by the components of the specific substrate
consumption vector at $(s_{1},s_{2})$. Thus, the construction of
growth isoclines and associated specific consumption vectors from
single-species data is essentially a method for characterizing the
behavior of the specific growth rate, $r_{i}^{g}(s_{1},s_{2})$, and
the specific substrate uptake rates, $r_{i1}^{s}(s_{1},s_{2}),r_{i2}^{s}(s_{1},s_{2})$.
By fixing the dilution rate in these experiments, we are choosing
particular paths on the $s_{1}s_{2}$-plane --- namely, the growth
isoclines at various $D$ --- along which the specific growth and
substrate uptake rates are determined. 

Now, there is nothing special about the paths corresponding to the
growth isoclines. We could just as well measure the specific growth
and consumption rates along any other path on the $s_{1}s_{2}$-plane.
The consumption curves described below represent another set of paths
for sampling the specific growth and substrate consumption rates in
single-species experiments.

\paragraph{The consumption curves and consumption vectors also capture the growth
and substrate uptake kinetics}

The consumption curve of the $i^{\mathrm{\mathrm{th}}}$ species at
feed concentrations, $s_{1}^{f},s_{2}^{f}$, denoted $\Phi_{i}(\mathbf{s^{f}})$,
is the locus of all steady state substrate concentrations obtained
when this species alone is grown on a mixture of $S_{1}$ and $S_{2}$
at \emph{fixed feed concentrations}, $s_{1}^{f},s_{2}^{f}$, but varying
dilution rates.

\begin{figure}[t]
\begin{center}\subfigure[]{\includegraphics[%
  width=7cm,
  height=5cm]{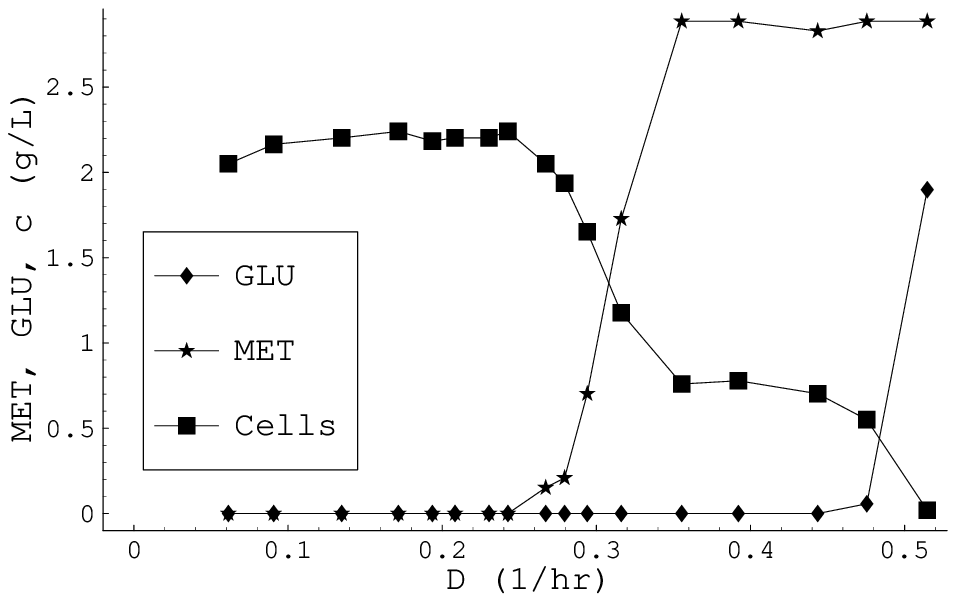}}\hspace{0.5in}\subfigure[]{\includegraphics[%
  width=7cm,
  height=5cm]{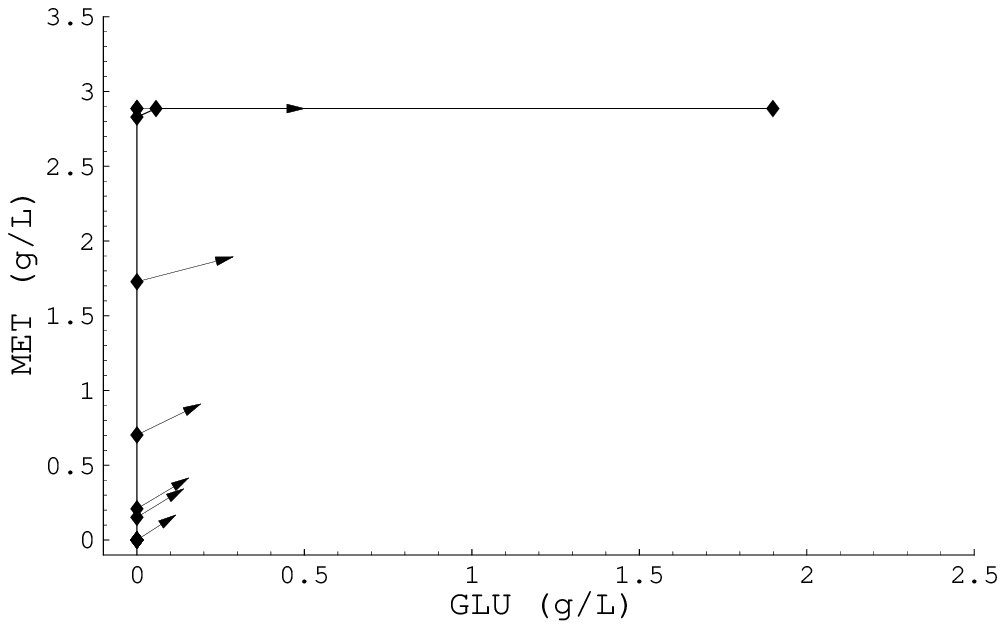}}\end{center}

\caption{\label{f:ConsumptionCurves1}Construction of the consumption curves:
(a)~The steady state substrate concentrations during growth of \emph{H.~polymorpha}
on a mixture of glucose and methanol at fixed feed concentrations
and varying $D$ (from~\citep{egli86b}). (b) The corresponding consumption
curve and associated specific substrate consumption vectors.}
\end{figure}

The upper panel of Figure~\ref{f:ConsumptionCurves1} illustrates
the construction of the consumption curve from single-species data.
Figure~\ref{f:ConsumptionCurves1}a shows the steady state glucose
and methanol concentrations obtained when \emph{H. polymorpha} is
grown on a mixture of glucose and methanol at fixed feed concentrations
and varying dilution rates. The corresponding consumption curve is
obtained by plotting the glucose and methanol concentrations at various
dilution rates on the plane of substrate concentrations (Figure~\ref{f:ConsumptionCurves1}b).
By appealing to~(\ref{eq:SpecificUptakeRates}), we can also calculate
the specific substrate uptake rates at each point of the consumption
curve, and represent them graphically as specific substrate consumption
vectors based at corresponding substrate concentrations (Figure~\ref{f:ConsumptionCurves1}b).

We gain insight into the biological significance of the consumption
curve by deriving its equation in terms of the model. To this end,
we observe that the steady states for single-species growth of the
$i^{{\rm th}}$ species must satisfy equations~(\ref{eq:ConsumptionCurve1}--\ref{eq:ConsumptionCurve2}).
These equations define the consumption curve, since eliminating $D$
from these equations yields the relation\begin{equation}
\frac{s_{2}^{f}-s_{2}}{s_{1}^{f}-s_{1}}=\frac{r_{i2}^{s}(s_{1},s_{2})}{r_{i1}^{s}(s_{1},s_{2})},\label{eq:MCconsumptionCurve}\end{equation}
which defines the locus of steady state substrate concentrations attained
at the fixed feed concentrations, $s_{1}^{f}$ and $s_{2}^{f}$. It
follows from the defining equations (\ref{eq:ConsumptionCurve1}--\ref{eq:ConsumptionCurve2})
that the consumption curve for the $i^{{\rm th}}$ species is the
locus of substrate concentrations at which the rates of supply of
both substrates are equal their consumption rates.

It is useful to obtain a more concise characterization of the consumption
curve by appealing to~(\ref{eq:MCconsumptionCurve}) rather than
the defining equations, (\ref{eq:ConsumptionCurve1}--\ref{eq:ConsumptionCurve2}).
To this end, define the \emph{substrate depletion vector}, $\triangle(\mathbf{s})\equiv(s_{1}^{f}-s_{1},s_{2}^{f}-s_{2})$,
at each pair of substrate concentrations. Then, (\ref{eq:MCconsumptionCurve})
says that the slopes of $\mathbf{r_{i}^{s}}(\mathbf{s})$ and $\triangle(\mathbf{s})$
are equal at every point of a consumption curve. Hence, the consumption
curve can also be viewed as the locus of all substrate concentrations
at which $\mathbf{r_{i}^{s}}(\mathbf{s})$ is parallel to $\triangle(\mathbf{s})$.
This property of consumption curves is evident in Figure~\ref{f:ConsumptionCurves1}b.

\begin{figure}[t]
\begin{center}\subfigure[]{\includegraphics[%
  width=7cm,
  height=4.5cm]{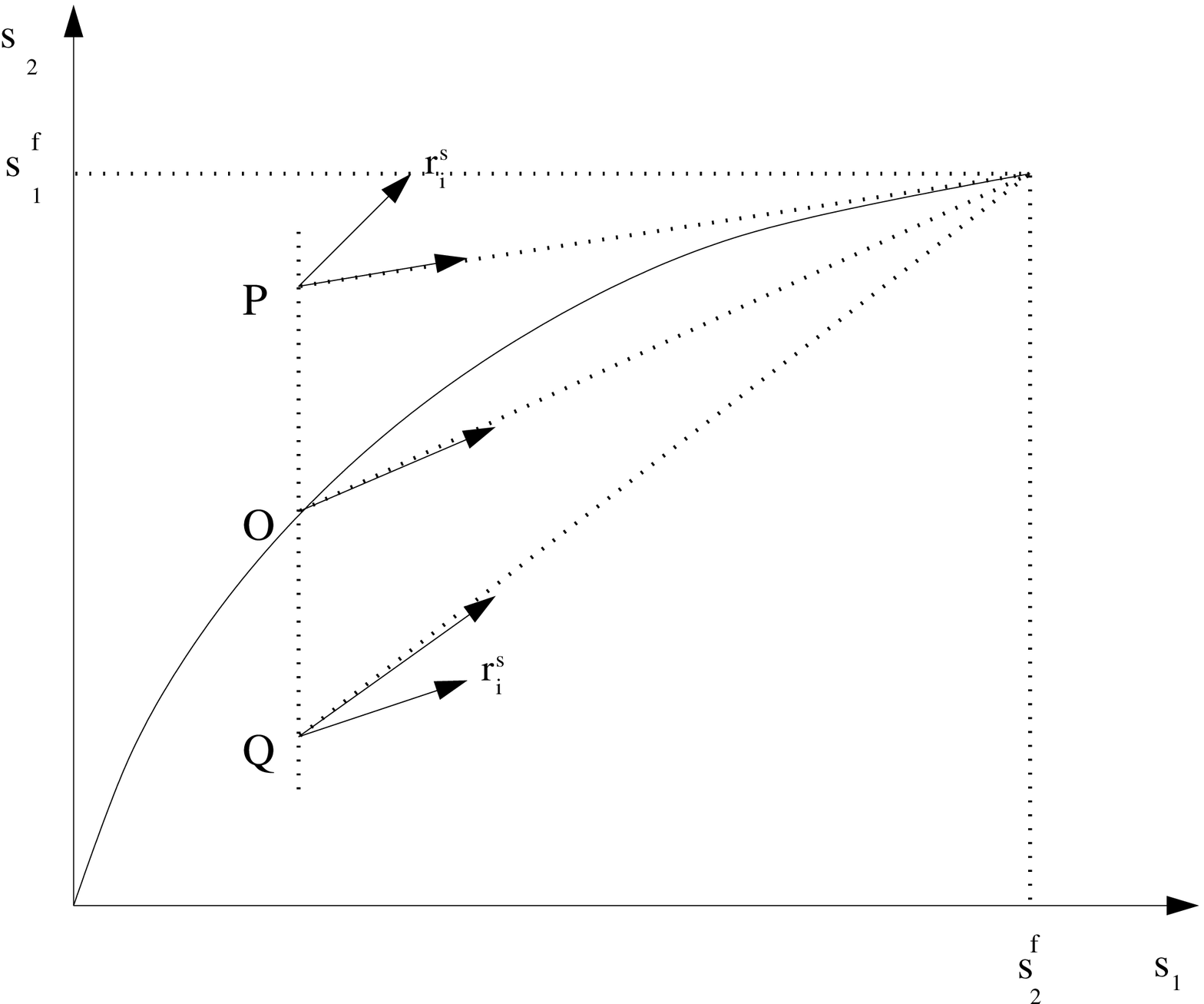}}\hspace{0.5in}\subfigure[]{\includegraphics[%
  width=7cm,
  height=4.5cm]{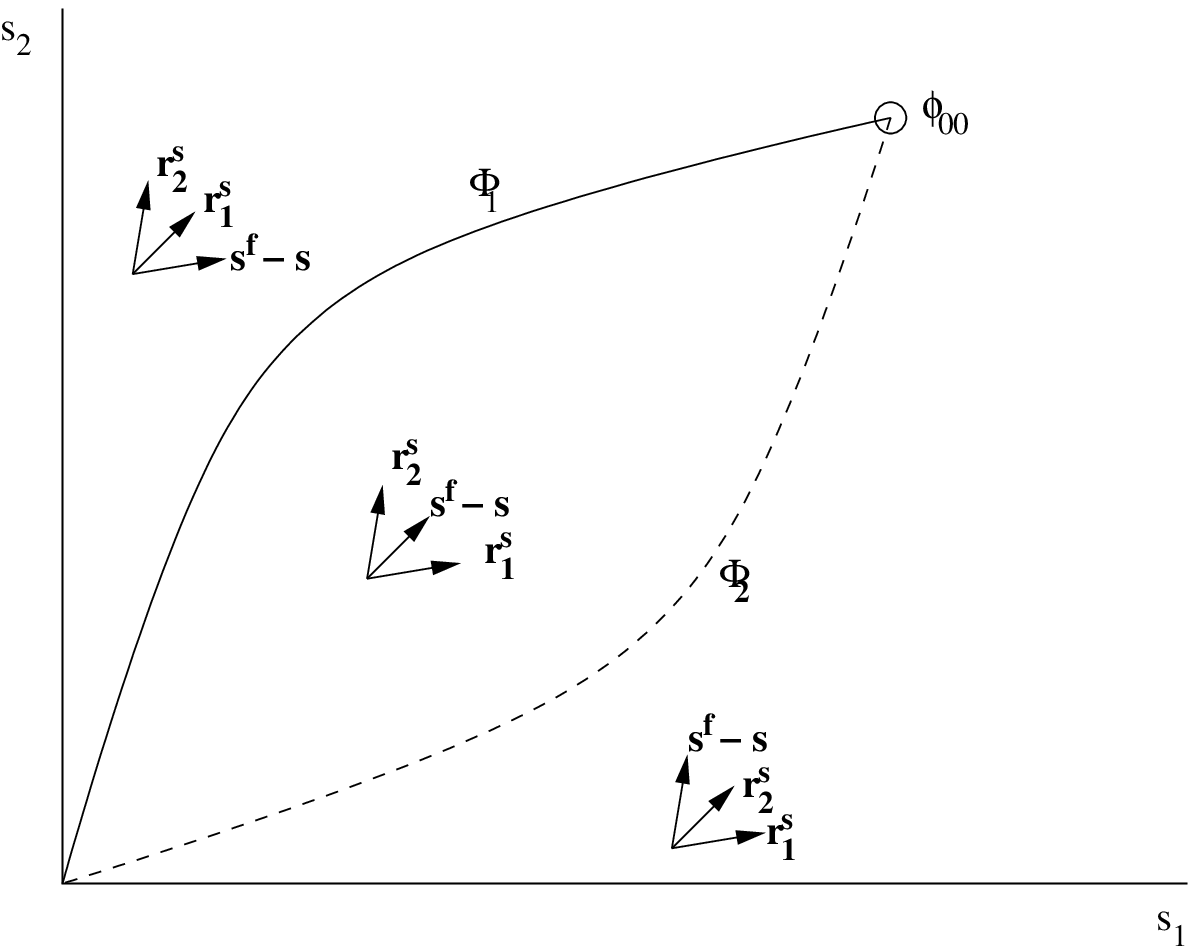}}\end{center}

\begin{center}\subfigure[]{\includegraphics[%
  width=7cm,
  height=4.5cm]{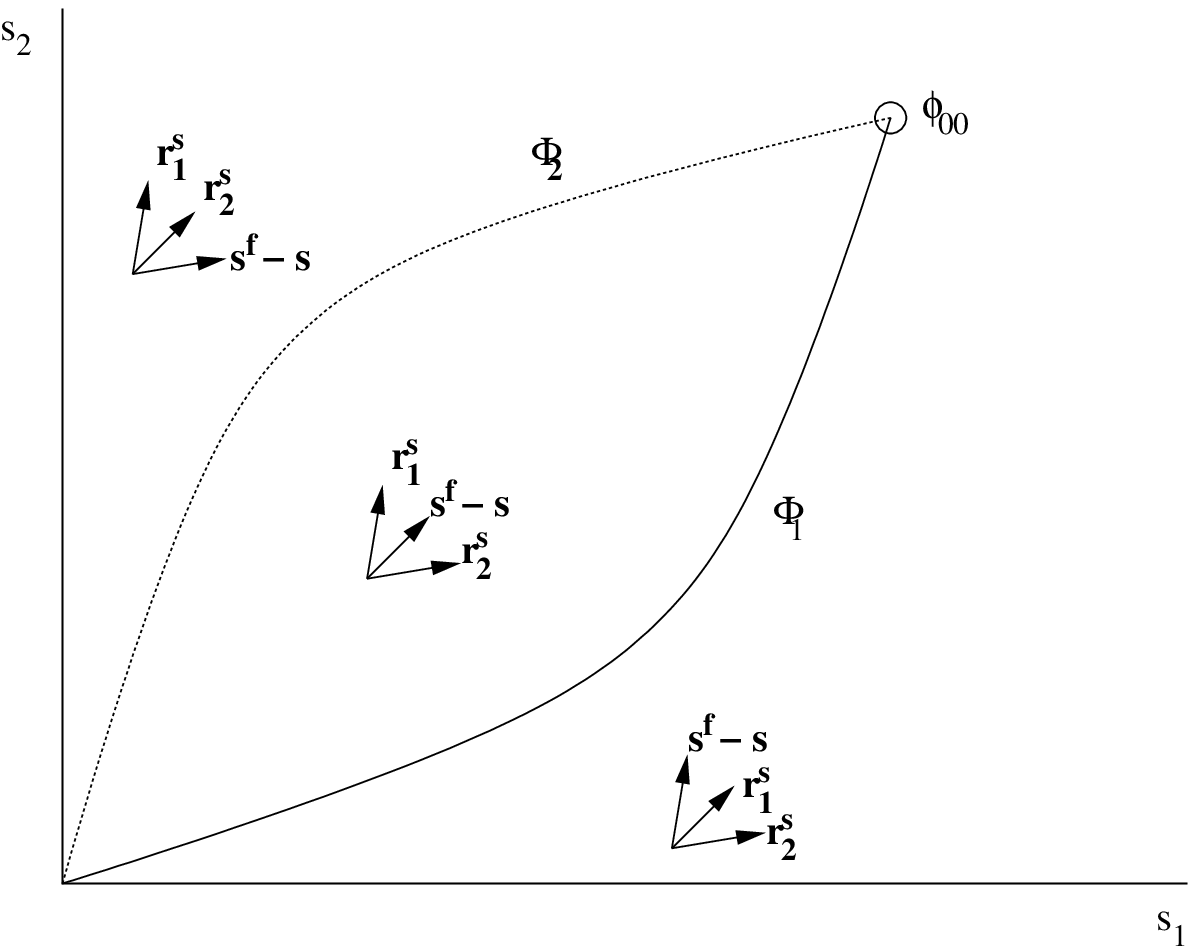}}\hspace{0.1in}\subfigure[]{\includegraphics[%
  width=7cm,
  height=4.5cm]{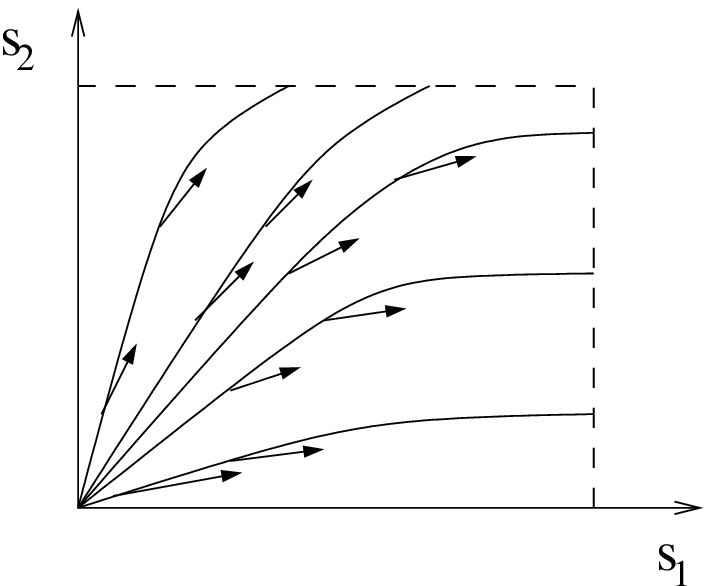}}\end{center}

\caption{\label{f:ConsumptionCurves2}Geometry of the consumption curves:
(a)~At each point of a consumption curve, the specific substrate
uptake vector, $\mathbf{r_{i}^{s}}(\mathbf{s})$, is parallel to the
substrate depletion vector, $\triangle(\mathbf{s})$. At substrate
concentrations above the consumption curve, $\mathbf{r_{i}^{s}}(\mathbf{s})$
is above $\triangle(\mathbf{s})$; at substrate concentrations below
the consumption curve, $\mathbf{r_{i}^{s}}(\mathbf{s})$ is below
$\triangle(\mathbf{s})$. (b, c)~The substrate depletion vector,
$\triangle(\mathbf{s})\equiv\mathbf{s^{f}-s}$, lies between the consumption
vectors, $\mathbf{r_{1}^{s}\mathnormal{(\mathbf{s})}}$ and $\mathbf{r_{2}^{s}}(\mathbf{s})$
at precisely those substrate concentrations that lie in the region
enclosed by the consumption curves for the two species denoted $\Phi_{1}$
and $\Phi_{2}$, respectively. The two figures depict the orientations
of the vectors when $\Phi_{1}$ lies above $\Phi_{2}$ and $\Phi_{2}$
lies above $\Phi_{1}$. (d)~A family of consumption curves (full
lines) and associated specific substrate consumption vectors constructed
by performing single-species experiments at various feed concentrations
(dashed lines).}
\end{figure}

At substrate concentrations above the consumption curve, the slope
of $\mathbf{r_{i}^{s}}(\mathbf{s})$ is higher than the slope of $\triangle(\mathbf{s})$
(Figure~\ref{f:ConsumptionCurves2}a). To see this, observe that
at a point $O$ on the consumption curve, $\mathbf{r_{i}^{s}}(\mathbf{s})$
is parallel to $\triangle(\mathbf{s})$. Now, suppose we move from
$O$ to $P$, so that $s_{1}$ is constant and $s_{2}$ increases.
Then the slope of $\triangle(\mathbf{s})\equiv(s_{1}^{f}-s_{1},s_{2}^{f}-s_{2})$
decreases and the slope of $\mathbf{r_{i}^{s}}(\mathbf{s})\equiv\left(r_{i1}^{s}(s_{1},s_{2}),r_{i2}^{s}(s_{1},s_{2})\right)$
increases (because $r_{i1}^{s}$ is a decreasing function of $s_{2}$,
and $r_{i2}^{s}$ is an increasing function of $s_{2}$). It follows
that at $P$, the slope of $\mathbf{r_{i}^{s}}(\mathbf{s})$ is higher
than the slope of $\triangle(\mathbf{s})$. If we move from $O$ to
$Q$ (Figure~\ref{f:ConsumptionCurves2}a), a similar argument shows
that at substrate concentrations below the consumption curve, the
slope of $\mathbf{r_{i}^{s}}(\mathbf{s})$ is lower than the slope
of $\triangle(\mathbf{s})$.

The foregoing fact immediately imply the following important result.
Given any feed concentrations, $\mathbf{s^{f}}$, and the corresponding
consumption curves, $\Phi_{1}(\mathbf{s^{f}})$, $\Phi_{2}(\mathbf{s^{f}})$,
for the two species, $\triangle(\mathbf{s})$ lies between $\mathbf{r_{i}^{s}}(\mathbf{s})$
and $\mathbf{r_{i}^{s}}(\mathbf{s})$ precisely when the substrate
concentrations lie between the two consumption curves (Figures~\ref{f:ConsumptionCurves2}b,c).
We shall appeal to this result later.

Figure~\ref{f:ConsumptionCurves1}b shows that as $D$ increases
from near-zero values to the washout dilution rate, the consumption
curve traverses an increasing path on the $s_{1}s_{2}$-plane that
starts at the origin, ($s_{1}=s_{2}=0)$, and terminates at the feed
point, $s_{1}=s_{1}^{f},s_{2}=s_{2}^{f}$. The same trend is predicted
by the model. Indeed, since the consumption curve satisfies the equation,
$r_{i1}^{s}(s_{1},s_{2})(s_{2}^{f}-s_{2})-r_{i2}^{s}(s_{1},s_{2})(s_{1}^{f}-s_{1})=0$,
it is clear that origin and the feed point lie on the consumption
curve. To see why the model predicts the increasing trend, note that
$\mathbf{r_{i}^{s}}(\mathbf{s})$ and $\triangle(\mathbf{s})$ must
be parallel along a consumption curve. This implies that as we move
along a consumption curve, both vectors must turn the same way (clockwise
or counterclockwise). However, one can check that along non-increasing
curves, $\mathbf{r_{i}^{s}}(\mathbf{s})$ and $\triangle(\mathbf{s})$
turn in opposite directions. It follows that consumption curves must
be increasing.

Figure~\ref{f:ConsumptionCurves2}d shows a hypothetical family of
consumption curves generated from single-species data. Just like the
family of growth isoclines, the family of consumption curves give
us complete information about the kinetics of growth and substrate
consumption. Indeed, each point on a consumption curve corresponds
to some dilution rate. Thus, given any point on the consumption curve,
the corresponding specific growth rate is equal to the dilution rate
at that point. The specific substrate uptake rates are given by the
specific substrate consumption vectors at that point. It is therefore
clear that Figures~\ref{f:GrowthIsoclines2}b and~\ref{f:ConsumptionCurves2}d
contain the same information. They merely represent two different
sets of paths for capturing this information.

\subsubsection{The two methods for characterizing the growth and substrate consumption
kinetics are equivalent}

The growth isoclines can be inferred from Figure~\ref{f:ConsumptionCurves2}d.
To generate the growth isocline at any given $D$, it suffices to
join the points on distinct consumption curves corresponding to this
$D$. Conversely, given the feed concentrations, we can generate the
corresponding consumption curve from Figure~\ref{f:GrowthIsoclines2}b.
For this, it suffices to scan each growth isocline until we find the
point $\mathbf{s}$ at which $\mathbf{r_{i}^{s}}(\mathbf{s})$ and
$\triangle(\mathbf{s})$ are parallel. Having determined such a point
on each growth isocline, we can join these points to generate the
consumption curve. 

Thus, we can perform single-species experiments to generate Figure~\ref{f:GrowthIsoclines2}b,
and infer the consumption curves from it. Alternatively, we can perform
single-species experiments to generate Figure~\ref{f:ConsumptionCurves2}d
and infer the growth isoclines. Both methods give us complete information
about the growth and substrate consumption kinetics of a given species
(although the latter approach seems easier than the former since it
involves nothing more than joining points corresponding to the same
$D$).

\subsection{Properties of mixed-culture steady states}

In what follows, it is assumed that by appealing to one of the two
methods described above, we have generated the growth isoclines and
consumption curves for both species at some dilution rate $D$ and
feed concentrations, $\mathbf{s^{f}}$. Our goal is to show that the
existence and stability of the mixed-culture steady states under these
conditions are completely determined by these curves. These results
have already been derived in terms of growth isoclines and consumption
vectors~\citep{leon75,tilman,tilman80}. Our goal is to show that
the consumption curves yield further insights into semitrivial and
nontrivial steady states.%
\footnote{Although the consumption curves provide no new information about the
trivial steady state, we have included a brief discussion of this
steady state for the sake of completeness.%
}

\begin{figure}[t]
\begin{center}\subfigure[Trivial steady state]{\includegraphics[%
  width=7cm,
  height=5cm]{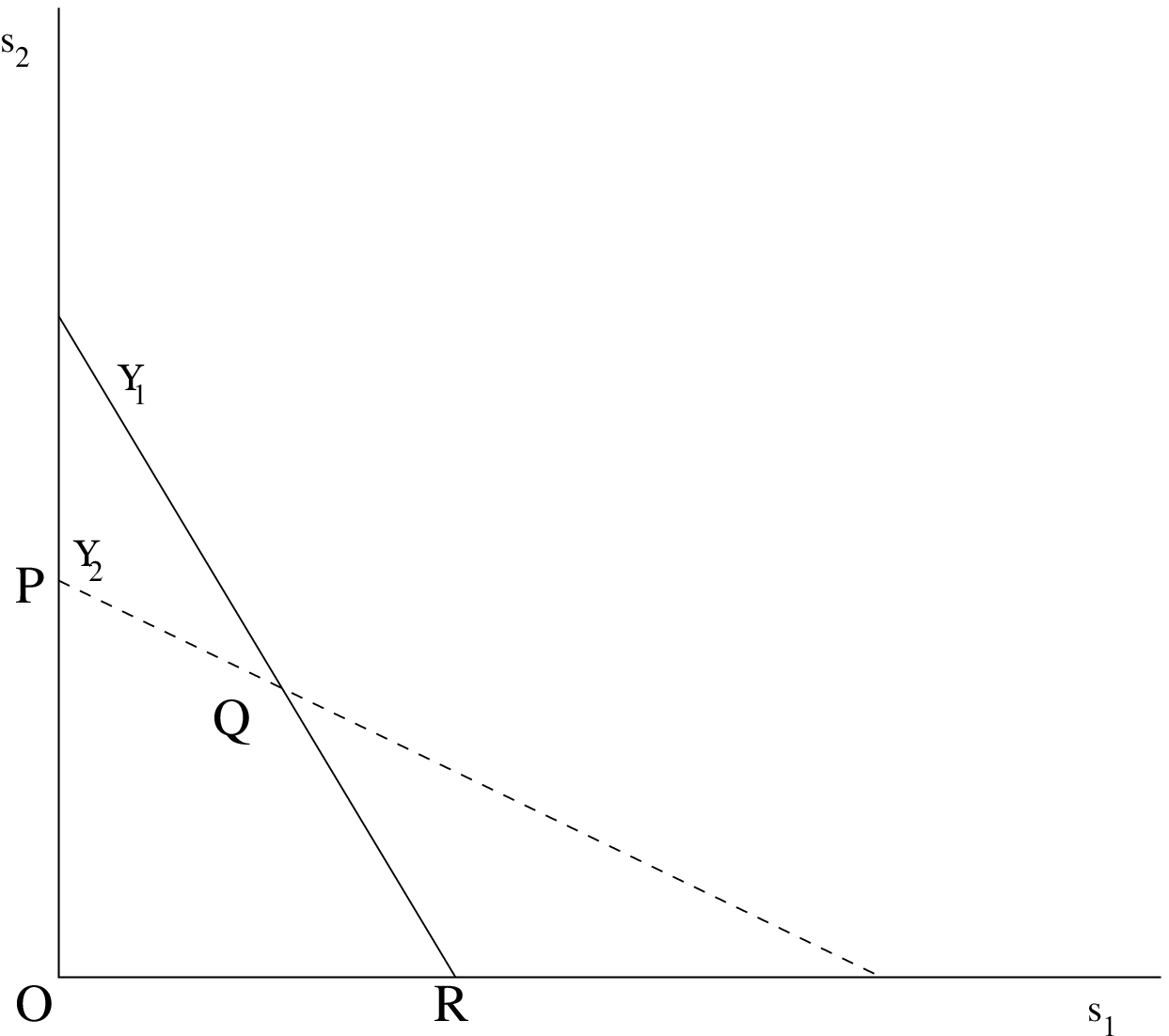}}\hspace{0.5in}\subfigure[Semitrivial steady state]{\includegraphics[%
  width=7cm,
  height=5cm]{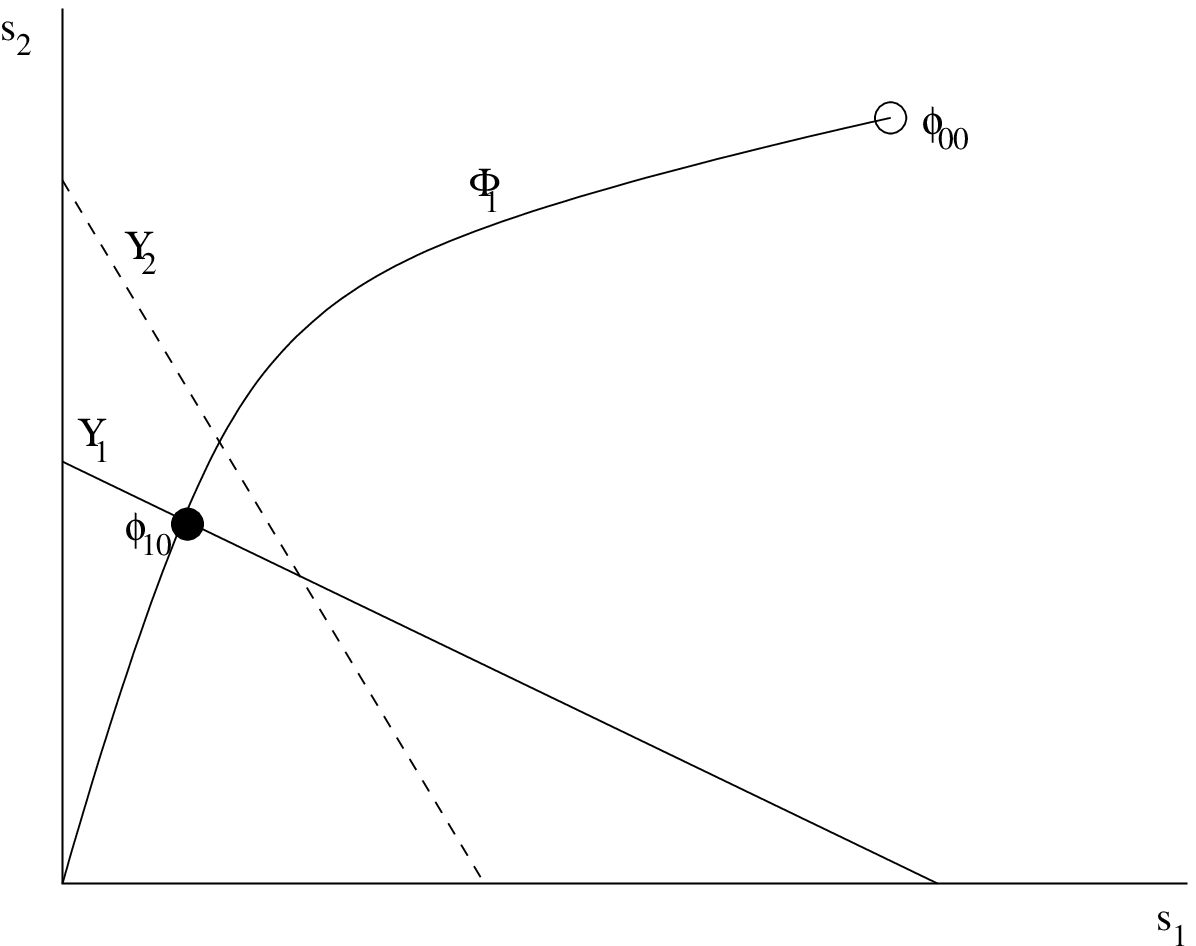}}\end{center}

\caption{\label{f:SemitrivialSS}Existence and stability of the trivial and
semitrivial steady states: (a) The trivial steady state exists for
all $D$ and $\mathbf{s^{f}}$. It is stable precisely when the feed
concentrations are in the region OPQR lying below the growth isoclines
for both species. (b) The semitrivial steady state, $\phi_{10}$,
lies at the intersection of the growth isocline, $\Upsilon_{1}(D)$,
and the consumption curve, $\Phi_{1}(\mathbf{s^{f}})$, for species
1. It is stable precisely when it lies below $\Upsilon_{2}(D)$, the
growth isocline for species two.}
\end{figure}

\paragraph{Trivial steady state}

At $\phi_{00}$, there are no cells in the reactor ($c_{1}=c_{2}=0)$
and the substrate concentrations are equal to the feed concentrations.
Thus, on the $s_{1}s_{2}$-plane, this steady state coincides with
the feed point, $\mathbf{s^{f}}$.

The trivial steady state always exists --- one can always arrange
to have a sterile chemostat at any dilution rate and feed concentrations.
However, from the mathematical point of view, this steady state is
not considered stable unless sterility is maintained even if the chemostat
is inoculated with both species. It is shown in Appendix~\ref{a:StabilitySS}
that the trivial steady state is stable precisely when $D>r_{1}^{g}(\mathbf{s^{f}}),r_{2}^{g}(\mathbf{s^{f}})$,
i.e., $D$ exceeds the maximum specific growth rates consistent with
the feed concentrations. At lower dilution rates, at least one of
the species will succeed in establishing itself in the chemostat.

In the case of weak mutual inhibition, the stability of the trivial
steady can be inferred by inspection of the two growth isoclines.
The trivial steady state is stable precisely when the feed concentrations
lie in the region OPQR lying below both growth isoclines (Figure~\ref{f:SemitrivialSS}a).
To see this, it suffices to recall that in the case of weak inhibition,
$r_{i}^{g}(\mathbf{s})<D$ for all $\mathbf{s}$ lying below the growth
isocline for the $i^{{\rm th}}$ species. Hence, $r_{1}^{g}(\mathbf{s_{f}}),r_{2}^{g}(\mathbf{s_{f}})<D$
for all $\mathbf{s^{f}}$ lying below both growth isoclines.

\paragraph{Semitrivial steady states}

We confine our attention to the semitrivial steady state, $\phi_{10}$,
at which only species~1 thrives in the chemostat ($c_{1}>0,c_{2}=0$).
The results for $\phi_{01}$ are analogous.

Given any $D$ and $\mathbf{s^{f}}$, the steady state substrate concentrations
at the corresponding semitrivial steady state(s) lie at the intersection
of the growth isocline, $\Upsilon_{1}(D)$, and the consumption curve,
$\Phi_{1}(\mathbf{s_{f}})$ (Figure~\ref{f:SemitrivialSS}b). To
see this, recall that 

\begin{enumerate}
\item The growth isocline, $\Upsilon_{1}(D)$, is the locus of all steady
state substrate concentrations attained when species 1 alone grows
at the dilution rate, $D$.
\item The consumption curve, $\Phi_{1}(\mathbf{s_{f}})$, is the locus of
all steady state substrate concentrations attained when species 1
alone grows at the feed concentrations, $\mathbf{s^{f}}$. 
\end{enumerate}
Hence, the steady state substrate concentrations during growth of
species~1 alone at the dilution rate, $D$, and feed concentrations,
$\mathbf{s^{f}}$, are given by the intersection point(s) of $\Upsilon_{1}(D)$
and $\Phi_{1}(\mathbf{s^{f}}$). It follows that this steady state
exists only if the feed concentrations lie above the growth isocline.
In the case of weak inhibition, there is at most one such steady state,
since $\Phi_{1}(\mathbf{s^{f}})$ is an increasing curve, and $\Upsilon_{1}(D)$
is a decreasing curve (Figure~\ref{f:GrowthIsoclines2}).

The mere existence of the semitrivial steady state, $\phi_{10}$,
does not imply that it will be observed in mixed-culture experiments.
It is experimentally observable if it is stable with respect to small
perturbations of the substrate concentrations and cell densities.
It is shown in Appendix~\ref{a:StabilitySS} that $\phi_{10}$ is
stable provided 

\begin{enumerate}
\item $\left.\nabla r_{1}^{g}\right|_{\phi_{10}}\cdot\left.\mathbf{t}_{1}\right|_{\phi_{10}}>0$,
where $\left.\mathbf{t}_{1}\right|_{\phi_{10}}$ denotes the tangent
to the consumption curve for species~1 at the substrate concentrations
corresponding to $\phi_{10}$, and oriented in the direction of increasing
substrate concentrations.
\item $\left.r_{2}^{g}\right|_{\phi_{10}}<D=\left.r_{1}^{g}\right|_{\phi_{10}}$,
where $\left.r_{i}^{g}\right|_{\phi_{10}}$ denotes the specific growth
rate of the $i^{{\rm th}}$ species at the steady state, $\phi_{10}$.
\end{enumerate}
These stability conditions have simple physical interpretations. The
first condition says that if the substrate concentrations are increased
ever so slightly along the consumption curve, the specific growth
rate must increase. The increase in the specific growth rate serves
to increase the cell density, thus negating the effect of the increased
substrate concentrations, and restoring the chemostat to its original
state.%
\footnote{This condition becomes more plausible if we observe that, in general,
the substrate concentrations are self-stabilizing. If the substrate
concentrations are increased slightly from their steady state values,
$ds_{j}/dt<0$ because $Ds_{j}$ and $r_{1j}^{s}(s_{1},s_{2})$ increase.
However, this is not the case when the substrate concentrations are
perturbed slightly along the tangent to the consumption curve. The
new substrate concentrations attained after such perturbations are,
to a first degree of approximation, still on the consumption curve,
so that $ds_{j}/dt=D(s_{j}^{f}-s_{j})-r_{1j}^{s}(s_{1},s_{2})c_{1}\approx0$.
Since the substrate concentrations are not self-stabilizing, the specific
growth rate, and hence, the cell density must increase to restore
the system to its original state.%
} The second condition says at $\phi_{10}$, the specific growth rate
of species~2 should be less than $D$, the specific growth rate at
which species~1 is already growing in the chemostat. This condition
ensures that if perturb the steady state by introducing a small inoculum
of species~2 into the chemostat, the chemostat will return to its
original state of single-species growth on species~1. 

The growth isoclines and consumption curves immediately tell us whether
these stability conditions will be satisfied in the mixed-culture
experiment. We can easily check if the first condition is satisfied,
since $\mathbf{t}_{1}$ is tangential to the consumption curve and
$\nabla r_{1}^{g}$ is perpendicular to the growth isocline. In the
case of weak mutual inhibition, the first condition is always satisfied
($\nabla r_{1}^{g}\cdot\mathbf{t}_{1}>0$), which reflects the fact
that under these conditions, $r_{1}^{g}$ is an increasing function
of $s_{1}$ and $s_{2}$. The second condition may or may not be satisfied.
To this end, recall that the $r_{2}^{g}<D$ precisely when the substrate
concentrations lie below the growth isocline for species 2. Thus,
$\phi_{10}$ is stable if the substrate concentrations at this steady
state lie below $\Upsilon_{2}(D)$ (Figure~\ref{f:SemitrivialSS}b);
it is unstable if the substrate concentrations at this steady state
lie above $\Upsilon_{2}(D)$. 

A similar argument for the other semitrivial steady state, $\phi_{01}$,
shows that the substrate concentrations at this steady state lie at
the intersection of $\Upsilon_{2}(D)$ and $\Phi_{2}(\mathbf{s^{f}})$.
This steady state is stable precisely when the corresponding substrate
concentrations lie below the growth isocline, $\Upsilon_{1}(D)$.

Hence, given the curves, $\Upsilon_{1}(D)$, $\Phi_{1}(\mathbf{s^{f}})$,
for species 1, and curves, $\Upsilon_{2}(D)$, $\Phi_{2}(\mathbf{s^{f}})$
for species 2, we can predict the existence and stability of $\phi_{10}$
and $\phi_{01}$ at $D$ and $\mathbf{s^{f}}$.

\paragraph{Nontrivial steady state}

It is not particularly surprising that the the growth isoclines and
consumption curves yield complete information about the existence
and stability of the semitrivial steady state, since these steady
states are identical to single-species steady states. However, it
is remarkable that they reveal the properties of the nontrivial steady
states, which have no counterpart in single-species cultures. 

It turns out that the substrate concentrations at the nontrivial steady
states are given by those intersection points of the two growth isoclines
that lie inside the region enclosed by the two consumption curves.
Thus, given the two growth isoclines, we can immediately identify
the substrate concentrations at all the coexistence steady states. 

To see why the nontrivial states can be obtained in this fashion,
observe that a nontrivial steady state satisfies the equations\begin{eqnarray}
0 & = & D(s_{j}^{f}-s_{j})-c_{1}r_{1j}^{s}(s_{1},s_{2})-c_{2}r_{2j}^{s}(s_{1},s_{2}),\quad j=1,2,\label{eq:MCss1}\\
0 & = & r_{i}^{g}(s_{1},s_{2})-D,\quad i=1,2,\label{eq:MCss2}\end{eqnarray}
It follows from~(\ref{eq:MCss2}) that the substrate concentrations
at a nontrivial steady states lie at the intersection points of the
two growth isoclines. However, all such substrate concentrations do
not necessarily correspond to coexistence steady states. To see this,
observe that the substrate concentrations must also satisfy equation~(\ref{eq:MCss1}),
which can be rewritten in the vectorial form\begin{equation}
c_{1}\mathbf{r_{1}^{s}}(\mathbf{s})+c_{2}\mathbf{r_{2}^{s}}(\mathbf{s})=D\triangle(\mathbf{s}).\label{eq:VectorialEquation}\end{equation}
 This equation has positive solutions, $c_{1},c_{2}>0$ precisely
when $\triangle(\mathbf{s})$ lies between $\mathbf{r_{1}^{s}}(\mathbf{s})$
and $\mathbf{r_{2}^{s}}(\mathbf{s})$. The latter is true precisely
when the substrate concentrations lie between the two consumption
curves (see Figures~\ref{f:ConsumptionCurves2}b,c). Thus, the substrate
concentrations at nontrivial steady states are given by only those
intersection points of the growth isoclines that lie within the region
enclosed by the two consumption curves. 

Given particular feed concentrations, $\mathbf{s^{f}}$, the consumption
curves for the two species delineate the substrate concentrations
that can be attained at coexistence steady states. In other words,
the coxistence steady state exists precisely when the two growth isoclines
intersect within the region enclosed by the consumption curves. Thus,
we have referred to the two consumption curves as the \emph{envelope
of coexistence}~\citep{pilyugin03a}.

\begin{figure}[t]
\begin{center}\subfigure[]{\includegraphics[%
  width=6.7cm,
  height=4.5cm]{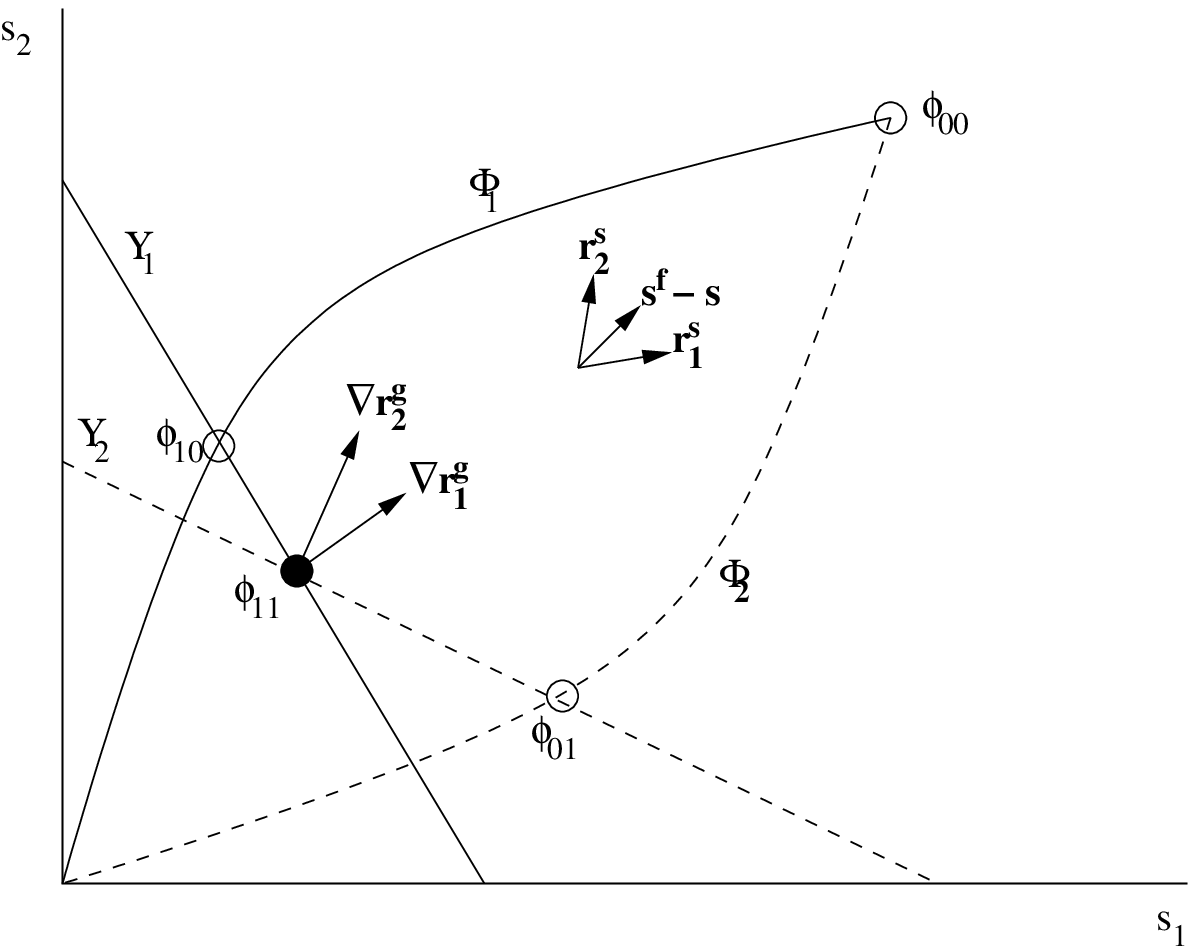}}\hspace*{0.2in}\subfigure[]{\includegraphics[%
  width=6.7cm,
  height=4.5cm]{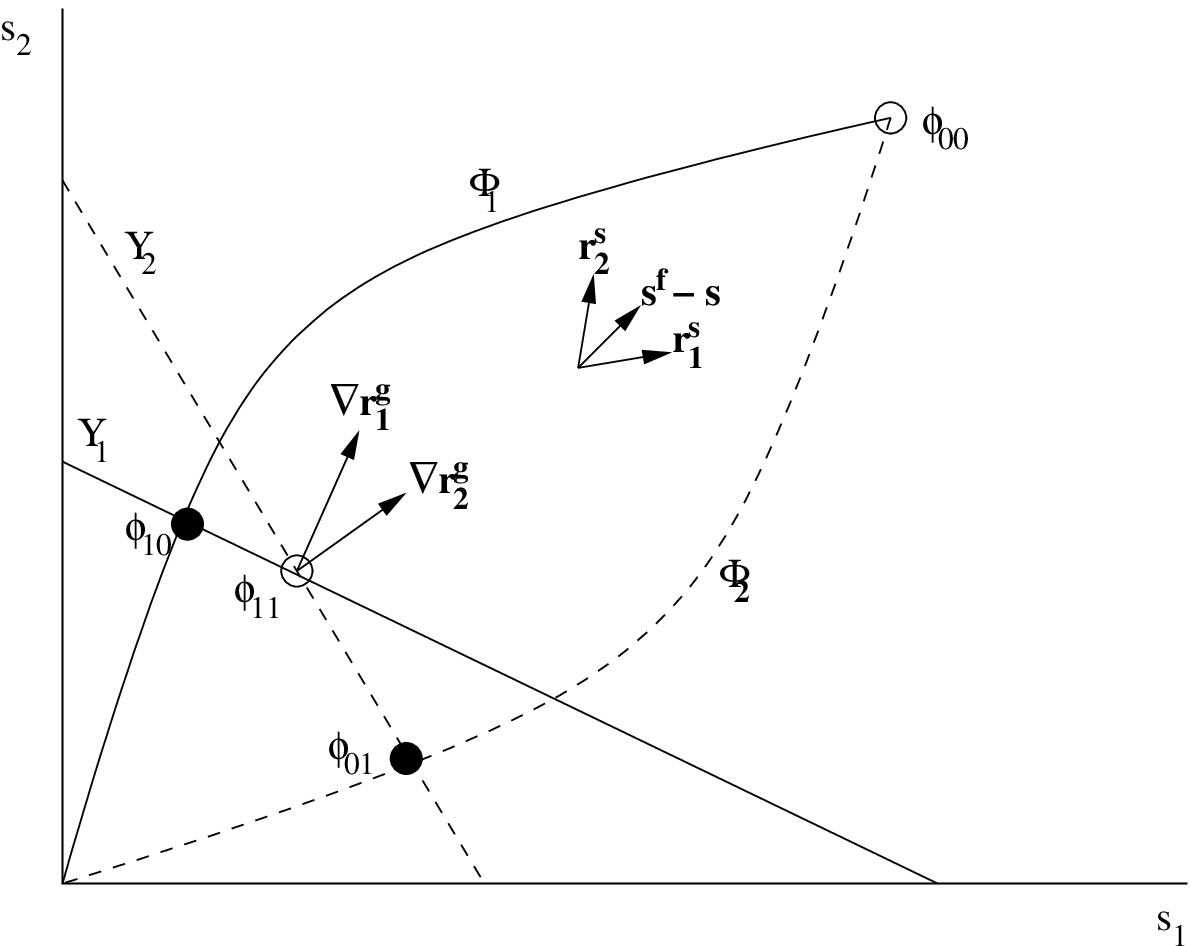}}\end{center}

\begin{center}\subfigure[]{\includegraphics[%
  width=6.7cm,
  height=4.5cm]{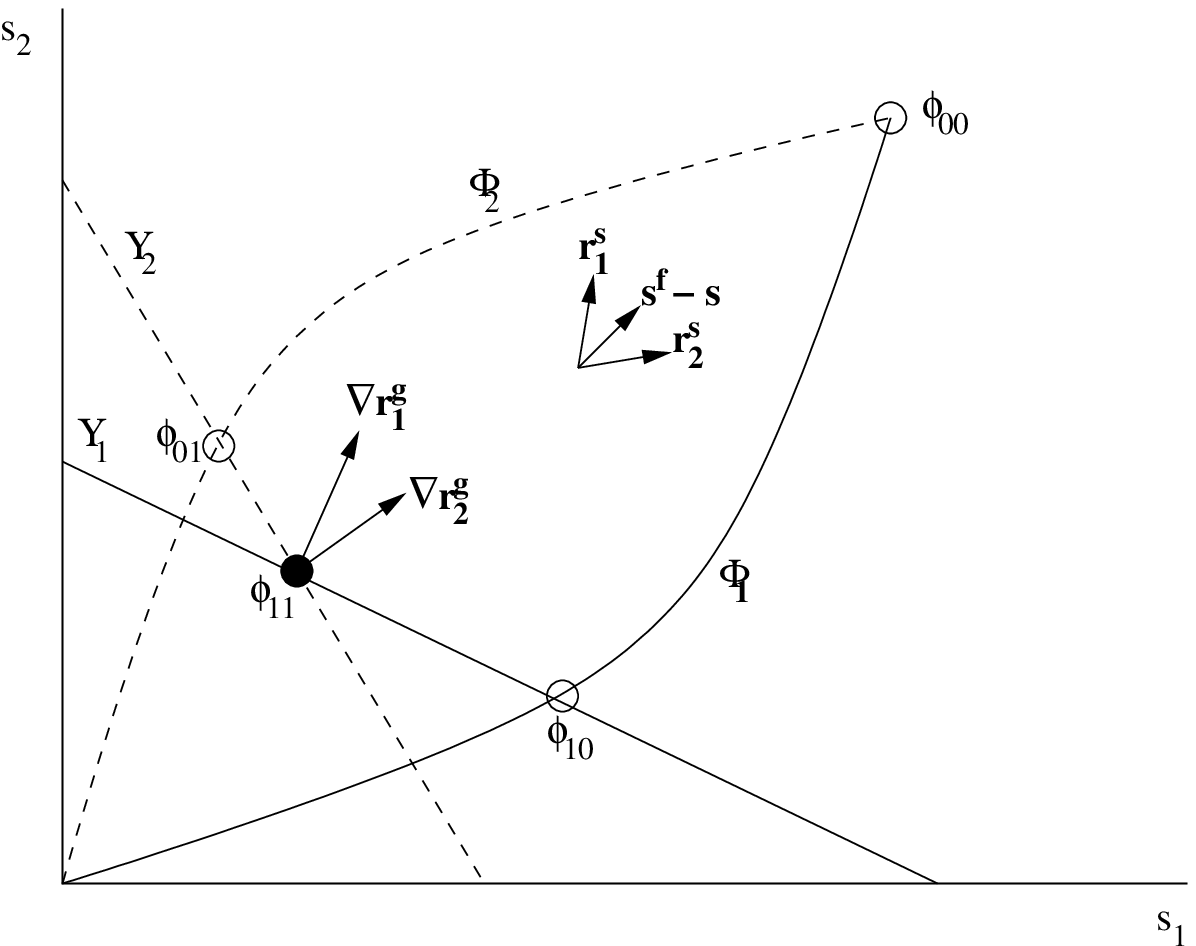}}\hspace*{0.2in}\subfigure[]{\includegraphics[%
  width=6.7cm,
  height=4.5cm]{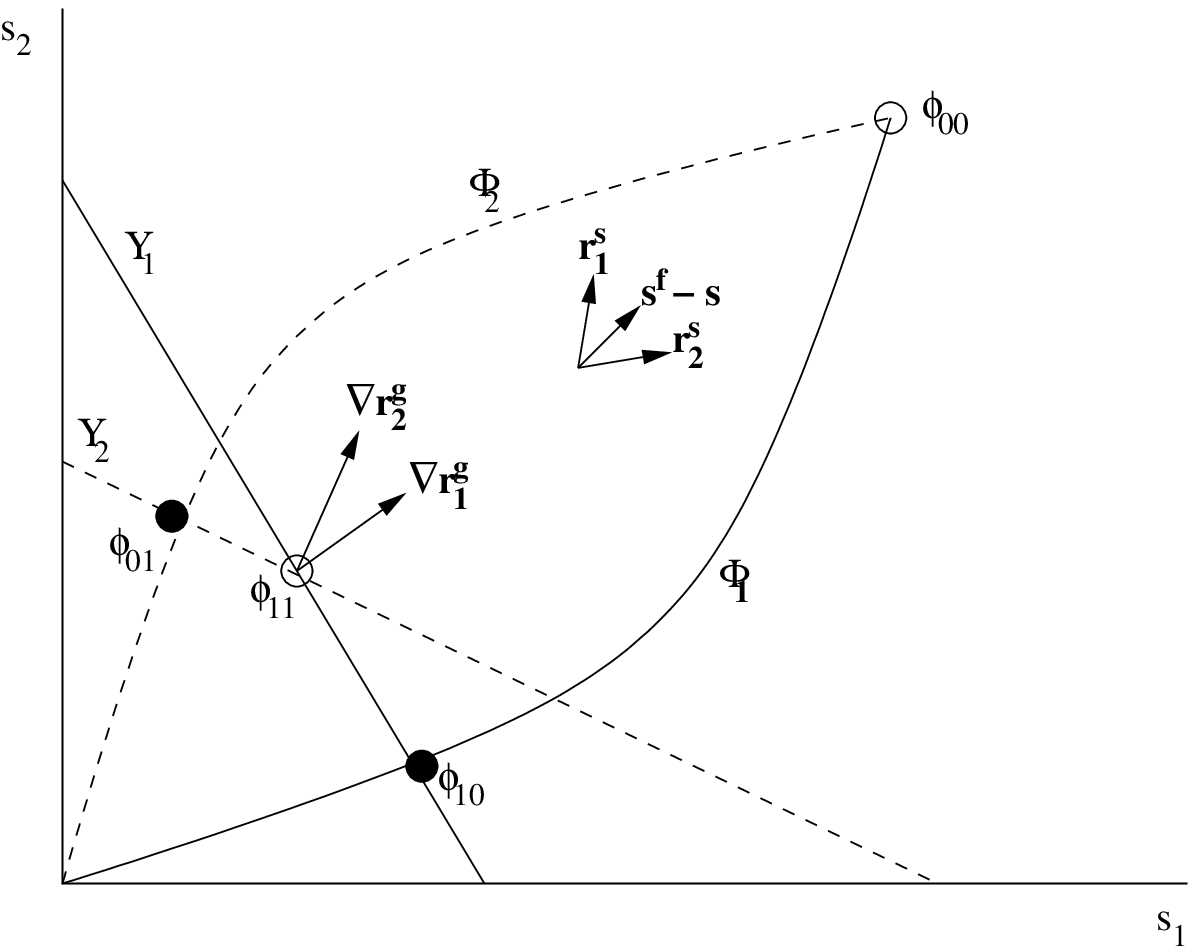}}\end{center}

\caption{\label{f:NontrivialStability}Inferring the stability of the mixed-culture
steady states at fixed $D$ and $\mathbf{s^{f}}$: The growth isocline
and consumption curve for species~1, denoted $\Phi_{1}$ and $\Upsilon_{1}$
are shown as full lines. The growth isocline and consumption curve
for species~2, denoted $\Phi_{2}$ and $\Upsilon_{2}$ are shown
as dashed lines. Stable and unstable steady states are shown as full
and open circles, respectively.}
\end{figure}

To determine whether the nontrivial steady state(s) are observable,
it is necessary to resolve the question of their stability. It is
shown in Appendix~\ref{a:StabilitySS} that a nontrivial steady state
is stable only if the two pairs of vectors, ($\mathbf{r_{1}^{s}},\mathbf{r_{2}^{s}}$)
and $(\nabla r_{1}^{g},\nabla r_{2}^{g})$ have the same \emph{orientation},
i.e.,

\begin{enumerate}
\item Either $\mathbf{r_{1}^{s}}$ lies above $\mathbf{r_{2}^{s}}$ and
$\nabla r_{1}^{g}$ lies above $\nabla r_{2}^{g}$ .
\item Or $\mathbf{r_{1}^{s}}$ lies below $\mathbf{r_{2}^{s}}$ and $\nabla r_{1}^{g}$
lies below $\nabla r_{2}^{g}$. 
\end{enumerate}
This stability criterion says that for a coexistence steady state
to be stable, it is necessary that each species consume more of the
substrate that limits its growth more~\citep{leon75,tilman80}. In
the first case, for instance, $\nabla r_{1}^{g}$ above $\nabla r_{2}^{g}$
means that growth of species~1 is more limited by $S_{2}$, and growth
of species~2 is more limited by $S_{1}$. The condition, $\mathbf{r_{1}^{s}}$
above $\mathbf{r_{2}^{s}}$ says that species~1 consumes more $S_{2}$,
and species~2 consumes more $S_{1}$. Thus, both species consume
more of the substrate that more strongly limits their growth ($S_{2}$
for species 1, and $S_{1}$ for species~2). The second case lends
itself to a similar interpretation.

The single-species data enables us to immediately test the validity
of this criterion. Indeed, since the growth isoclines are known, we
can find the orientations of $\nabla r_{1}^{g}$ and $\nabla r_{2}^{g}$
at a coexistence steady state --- they are perpendicular to the growth
isoclines at the coexistence steady state. Given the consumption curves,
we can also infer the orientations of $\mathbf{r_{1}^{s}}$ and $\mathbf{r_{2}^{s}}$
at any coexistence steady state (see Figure~\ref{f:ConsumptionCurves2}b,c).
In fact, since we know the consumption vectors at various points on
both growth isoclines, we calculate $\mathbf{r_{1}^{s}}$ and $\mathbf{r_{2}^{s}}$
at any coexistence steady state. Given these vectors, we can predict
the cell densities of both species at the coexistence steady state
by solving equation~(\ref{eq:VectorialEquation}).

\subsection{Applications of the theory}

\begin{figure}[t]
\begin{center}\subfigure[]{\includegraphics[%
  width=7cm,
  height=5cm]{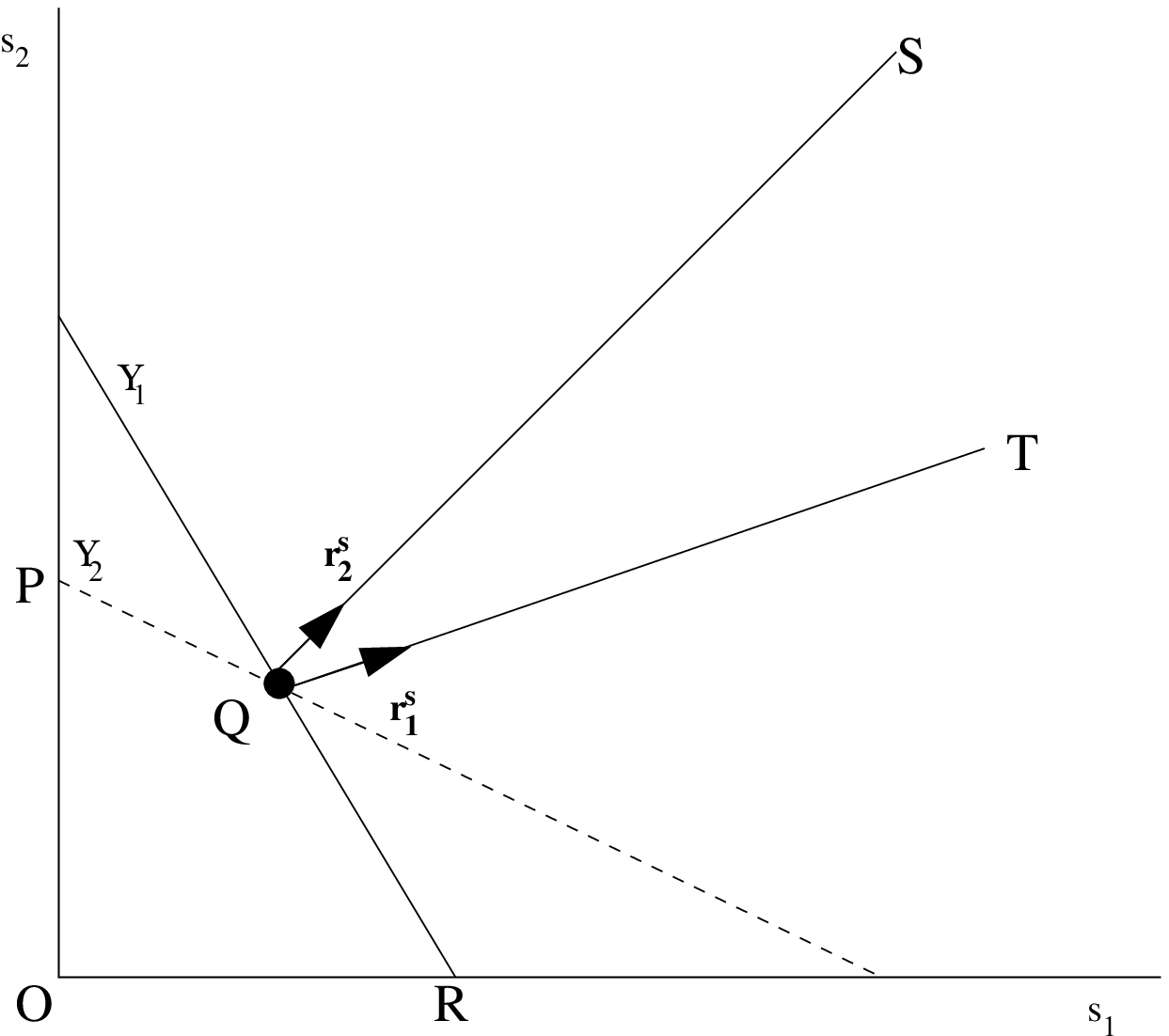}}\hspace{0.5in}\subfigure[]{\includegraphics[%
  width=7cm,
  height=5cm]{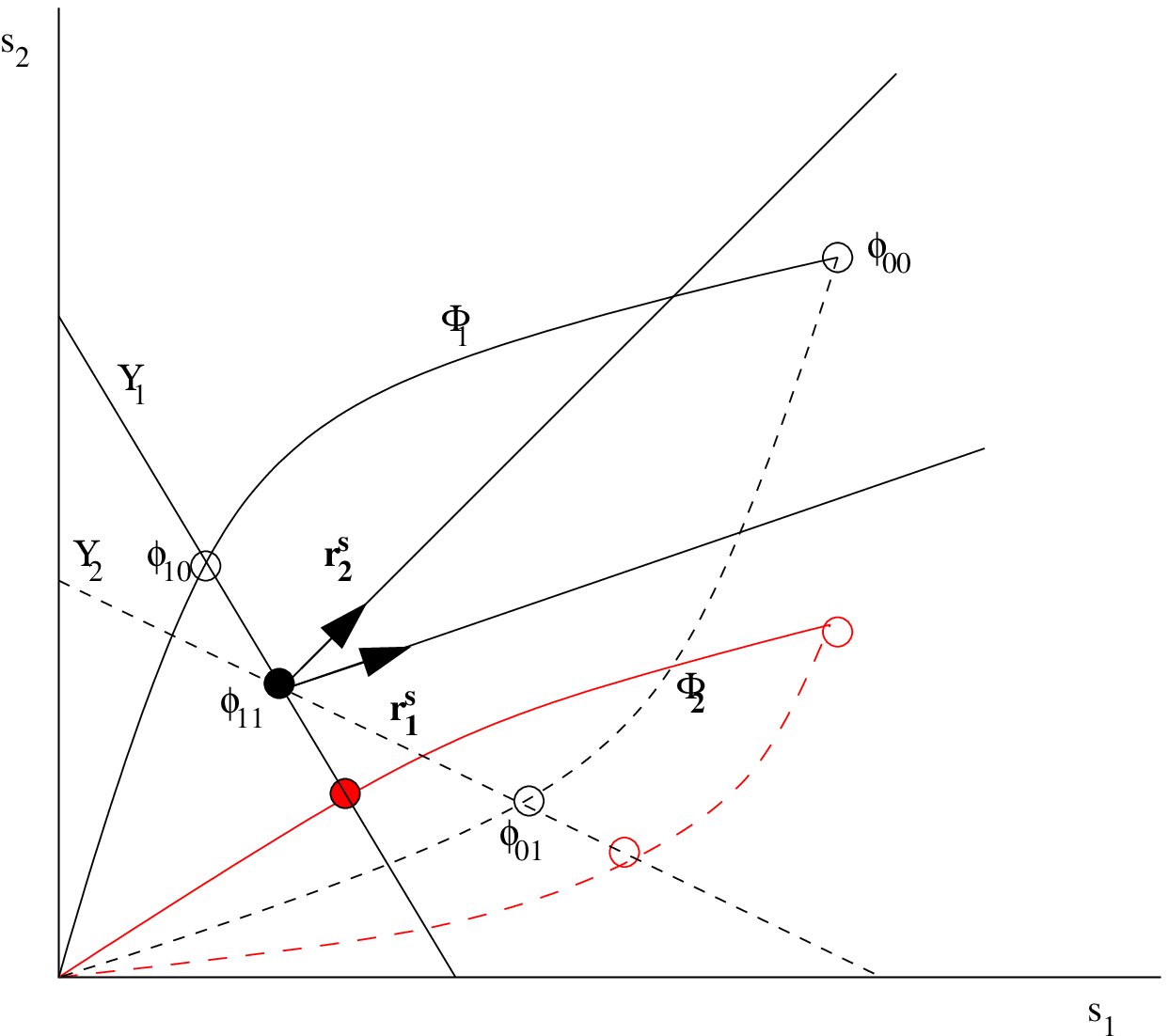}}\end{center}

\caption{\label{f:ResourceRatio}Inferring the stability of the mixed-culture
steady states at fixed $D$ and varying $\mathbf{s^{f}}$: (a)~The
properties of the steady states at various feed concentrations can
be inferred by drawing the cone, SQT, generated by the specific substrate
consumption vectors, $\mathbf{r_{1}^{s}}$ and $\mathbf{r_{2}^{s}}$,
at the coexistence steady state. Coexistence is stable if the feed
concentrations lie in the cone, SQT. Both species wash out if the
feed concentrations lie in the region, OPQR. Only one of the species
survives if the feed concentrations lie in the region, PQS, or RQT.
(b)~Explanation of the results in (a) based on the consumption curves.
If $s_{2}^{f}$ is decreased while $D$ and $s_{1}^{f}$ are fixed,
the consumption curves for both species move down. Consequently, the
coexistence steady state, $\phi_{11}$, is pushed out of the envelope
of coexistence. At these new feed concentrations, only species~1
survive will survive in the chemostat.}
\end{figure}

In what follows, we apply the foregoing results to various hypothetical
cases. The conclusions reached will form the basis for the experiments
described in the next section.

\subsubsection{Fixed dilution rate and feed concentrations}

We begin by considering the implications of the theory when the dilution
rate rate and feed concentrations are fixed. Specifically, we assume
that the growth isoclines and consumption curves at some dilution
rate, $D$, and feed concentrations, $\mathbf{s^{f}}$, have been
constructed from single-species experiments. We show that we can the
properties of all the mixed-culture steady states that would be obtained
at this $D$ and $\mathbf{s^{f}}$.

Figure~\ref{f:NontrivialStability} shows 4 hypothetical arrangements
of the growth isoclines and consumption curves for the two species.
We begin by the applying the results to the arrangement shown in Figure~\ref{f:NontrivialStability}a.
In this case

\begin{enumerate}
\item The semitrivial steady state, $\phi_{10}$, lies at the intersection
of the growth isocline and the consumption curve for species~1 ($\Upsilon_{1}(D)$
and $\Phi_{1}(\mathbf{s^{f}})$, respectively). It is unstable because
it lies above $\Upsilon_{2}(D)$, the growth isocline for species~2.
\item The semitrivial steady state, $\phi_{01}$, lies at the intersection
of the growth isocline and the consumption curve for species~2 ($\Upsilon_{2}(D)$
and $\Phi_{2}(\mathbf{s^{f}})$, respectively). It is unstable because
it lies above $\Upsilon_{1}(D)$, the growth isocline for species~1.
\item The nontrivial steady state, $\phi_{11}$, lies at the intersection
of the two growth isoclines. It is a legitimate nontrivial steady
state because the corresponding substrate concentrations lie between
the two consumption curves. It is likely to be stable since $\mathbf{r_{1}^{s}}$
lies below $\mathbf{r_{2}^{s}}$ and $\nabla r_{1}^{g}$ lies below
$\nabla r_{2}^{g}$.
\end{enumerate}
If the growth isoclines and consumption curves have the disposition
shown in Figure~\ref{f:NontrivialStability}b

\begin{enumerate}
\item The semitrivial steady state, $\phi_{10}$, which lies at the intersection
of $\Upsilon_{1}(D)$ and $\Phi_{1}(\mathbf{s^{f}})$, is stable because
it lies below $\Upsilon_{2}(D)$, the growth isocline for species~2.
\item The semitrivial steady state, $\phi_{01}$, which lies at the intersection
of $\Upsilon_{2}(D)$ and $\Phi_{2}(\mathbf{s^{f}})$, is stable because
it lies below $\Upsilon_{1}(D)$, the growth isocline for species~1.
\item The nontrivial steady state, $\phi_{11}$, which lies at the intersection
of the two growth isoclines, is certainly unstable since $\mathbf{r_{1}^{s}}$
lies below $\mathbf{r_{2}^{s}}$, but $\nabla r_{1}^{g}$ lies above
$\nabla r_{2}^{g}$.
\end{enumerate}
The remaining two cases in Figure~\ref{f:NontrivialStability} can
be understood by similar arguments.

\subsubsection{Fixed dilution rate and varying feed concentrations}

Next, we consider the behavior in response to varying feed concentrations.
Specifically, we assume that the chemostat is at a (stable) coexistence
steady state at some dilution rate, $D$, and feed concentrations,
$\mathbf{s^{f}}$, and we ask: How does the chemostat respond if the
dilution is fixed, and the feed concentrations are changed.

Tilman obtained the complete answer to this question~\citep{tilman80,tilman}.
Figure~\ref{f:ResourceRatio}a shows the behavior that would be observed
at all possible feed concentrations. Since the dilution rate is fixed,
the growth isoclines, and hence, the substrate concentrations at the
coexistence steady state remain unchanged. Tilman showed that 

\begin{enumerate}
\item Both species coexist precisely when the feed concentrations lie in
the cone, SQT, generated by the specific substrate consumption vectors
at the coexistence steady state. This follows immediately from (\ref{eq:VectorialEquation}),
which says that coexistence steady states exist precisely when $\triangle(\mathbf{s})$
lies between $\mathbf{r_{1}^{s}}$ and $\mathbf{r_{2}^{s}}$. 
\item Both species wash out when the feed concentrations lie in the region,
OPQR. 
\item Only one of the species thrives when the feed concentration lie in
the regions above PQS or below RQT.
\end{enumerate}
It follows from this picture that if the ratio of the feed concentrations,
$s_{2}^{f}/s_{1}^{f}$, is too high or too low, one of the species
will be rendered extinct. This result, often referred to as the \emph{resource-ratio
hypothesis}, is an important consequence of the theory. It suggests
an explanation for the \emph{paradox of enrichment} --- enrichment
of an ecosystem with only one of nutrients reduces its biodiversity.

By appealing to consumption curves, we can see exactly why the coexistence
steady state ceases to exist if the feed concentrations lie outside
the cone, SQT. Figure~\ref{f:ResourceRatio}b shows the manner in
which the consumption curves are shifted when $s_{2}^{f}$ is decreased
at fixed $s_{1}^{f}$ (the original and new consumption curves are
shown as black and red curves, respectively). At sufficiently small
$s_{2}^{f}$, the growth isoclines intersect outside the envelope
of coexistence; hence, the coexistence steady state ceases to exist.
Under these conditions, species~1 becomes dominant, since the other
semitrivial steady state, $\phi_{01}$, is unstable.

\subsection{The case of strong mutual inhibition}

The experimental data suggests that in some instances, the assumption
of weak inhibition ($\partial r_{i}^{g}/\partial s_{1},\partial r_{i}^{g}/\partial s_{2}>0)$
is violated. For instance, the specific growth rates of \emph{H.~polymorpha}
on glucose, xylose, and glycerol are 0.61, 0.175, and 0.27~1/hr,
respectively~\citep{Brinkmann92}. During mixed-substrate growth
on glucose + xylose and glucose + glycerol, the specific growth rates
are 0.36~1/hr and 0.37~1/hr, respectively. Thus, addition of xylose
or glycerol to a culture growing on glucose decreases the growth rate,
whereas addition of glucose to a culture growing on xylose or glycerol
increases the specific growth rate. Likewise, addition of 3-phenylpropionic
acid to a culture of \emph{E.~coli} ML308 growing on glucose decreases
the growth rate, whereas addition of glucose to a culture growing
on 3-phenylpropionic acid increases the growth rate~\citep{Kovarova96}.
It follows that for certain substrate concentrations, $\partial r_{i}^{g}/\partial s_{1},\partial r_{i}^{g}/\partial s_{2}$
have opposite signs, and the growth isoclines may not be monotonically
decreasing.

When the growth isoclines are not monotonic, there may be multiple
semitrivial steady states. Moreover, some of these steady states may
be unstable because $\nabla r_{1}^{g}.\mathbf{t}_{1}$ is negative
(a condition that is impossible in the case of weak inhibition). This
instability arises because the specific growth rate is a decreasing
function of the substrate concentrations along certain directions
in the neighborhood of the steady state. Perturbations that increase
the substrate concentrations along such directions decrease the specific
growth rate, which destabilizes the system by reducing the cell density
in the chemostat.

Thus, in the case of strong mutual inhibition, the consumption curves,
which provide the vector, $\mathbf{t_{1}}$, are crucial for determining
the stability of the semitrivial steady states.

\section{\label{s:Discussion}Discussion}

We have shown above that a complete theory of mixed-culture growth
on substitutable substrates can be developed without making specific
assumptions about the kinetic expressions. Since the theory appeals
directly to the data, the assumptions required for predicting mixed-culture
growth are relatively weak. We have also shown that the consumption
curves can be inferred from the single-species data, and that they
provide additional insights into the existence and stability of the
mixed-culture steady states. In what follows, we describe experiments
that may be designed to test the hypotheses and predictions of the
theory.

\subsection{Experiments for testing the validity of the model hypotheses}

The first hypothesis (\ref{eq:HypothesisUniqueness}) states that
the specific growth and uptake rates depend only on the substrate
concentrations --- they are independent of the cell densities. If
this hypothesis is true, then 

\begin{enumerate}
\item Equations (\ref{eq:MCgrowthIsocline}) and (\ref{eq:SpecificUptakeRates})
imply that the growth isocline and the associated consumption vectors
are completely determined by the identity of the species and the dilution
rate --- they are independent of the feed concentrations. For example,
the growth isoclines and specific substrate consumption vectors shown
in Figure~\ref{f:GrowthIsoclines2}b were generated by varying the
feed composition at fixed $D=0.3$~1/hr and $s_{1}^{f}+s_{2}^{f}=100$~mg/L.
If the specific growth and consumption rates are truly independent
of the cell density, the very same growth isocline and specific substrate
consumption vectors would be obtained if the experiments were performed
at the same $D$, but different total feed concentration (say, 200~mg/L).
\item Equation (\ref{eq:MCss2}) implies that the steady state concentrations
of a coexistence steady state are independent of the feed concentrations.
To test this hypothesis, suppose we have a coexistence steady state
at some dilution rate and feed concentrations. If our hypothesis is
correct, the steady state substrate concentrations will not change
even if we change the feed concentrations.
\end{enumerate}
These experiments are analogous to those performed for verification
of the Monod model for single-substrate, single-species growth. The
Monod model predicts that the steady state substrate concentration
depends only on the dilution rate --- it is independent of the feed
concentrations. Experiments have shown that this is indeed the case.

The second hypothesis is that each substrate promotes its own own
uptake and inhibits the uptake of the other substrate (\ref{eq:HypothesisMChomo}--\ref{eq:HypothesisMChetero}).
This hypothesis implies that the consumption curve is a strictly increasing.
Thus, the experimental consumption curves provide an indirect test
of this hypothesis. It can also be tested directly by transferring
a sample of the steady state single-species culture in the chemostat
to a shake flask, spiking the sample with a small amount of a substrate,
and measuring the uptake rates of both substrates. Such experiments
have been performed by Lendenmann \& Egli~\citep{lendenmann95}.

The third and final hypothesis is that the yields are constant. An
indirect test follows from the fact that if this hypothesis is true,
the steady state cell density in single-species experiments is given
by\[
c_{i}=Y_{i1}(s_{1}^{f}-s_{1})+Y_{i2}(s_{2}^{f}-s_{2})\]
where $Y_{i1},Y_{i2}$ are the yields of the $i^{{\rm th}}$ species
during single-substrate growth on $S_{1}$ and $S_{2}$. If the observed
cell densities do not agree with the cell densities calculated from
this expression, the hypothesis cannot be true. Direct tests of this
hypothesis can be performed with radioactively labeled substrates
(see~\citep{egli82a} for an example)

\subsection{Experiments for testing the validity of the model predictions}

Few predictions of resource-based theory have been tested rigorously
(see~\citep{Miller2005} for a comprehensive review). Figures~\ref{f:NontrivialStability}
and~\ref{f:ResourceRatio} illustrate some predictions of the theory.
Specifically, the model predicts

\begin{enumerate}
\item The dilution rates and feed concentrations at which coexistence will
be observed. 
\item The substrate concentrations and cell densities attained at any coexistence
state.
\item The manner in which the chemostat will respond if the feed concentrations
are varied at a fixed dilution rate.
\end{enumerate}
All these predictions can be tested experimentally. Insofar as substitutable
substrates are concerned, the last prediction was tested by Rothhaupt
with a system in which two rotifers competed for growth on two algal
species~\citep{Rothhaupt1988}. However, none of these predictions
have been tested in microbial systems.

\section{\label{s:Conclusions}Conclusions}

We extended the theory of mixed-culture growth on mixtures of substrates
in two directions. Specifically, we showed that

\begin{enumerate}
\item The single-species data completely determines the behavior of the
mixed culture. It is not necessary to make specific assumptions about
the kinetics of growth and substrate consumption. Relatively weak
assumptions about the qualitative properties of growth and substrate
consumption are sufficient for inferring the properties of the mixed
culture.
\item There is nothing special about the growth isoclines. They represent
a set of paths on the $s_{1}s_{2}$-plane along which one can acquire
information about the growth and substrate consumption rates. The
same information can be obtained by choosing another set of paths,
namely, the consumption curves. Importantly, the consumption curves
can be deduced from the growth isoclines and consumption vectors;
conversely, the growth isoclines can be inferred from the consumption
curves. The two sets of curves are therefore {}``dual'' to each
other.
\item Consideration of the growth isoclines together with the consumption
curves yields deeper insights into the model. Specifically, we showed
that the consumption curves delineate all possible substrate concentrations
attained by coexistence steady state, enable us to define stability
criteria for systems in which the growth isoclines are non-monotonic,
and elucidate the bifurcations in response to varying dilution rates
and feed concentrations..
\end{enumerate}
Finally, we discussed experiments that can be used to test the validity
of the model hypotheses and predictions.

\smallskip{}
\noindent \textbf{Acknowledgements:} We thank Prof.~Robert Holt (Zoology,
University of Florida) for comments on an early draft of this manuscript.

\bibliographystyle{plainnat}
\bibliography{growthkinetics}

\appendix

\section{Stability of the steady states \label{a:StabilitySS}}

\subsection{Trivial steady state}

To determine the stability of $\phi_{00}$, consider the Jacobian
\[
J(\phi_{00})=\left(\begin{array}{cccc}
-D & 0 & -r_{11}^{s} & -r_{21}^{s}\\
0 & -D & -r_{12}^{s} & -r_{22}^{s}\\
0 & 0 & r_{1}^{g}-D & 0\\
0 & 0 & 0 & r_{2}^{g}-D\end{array}\right).\]
 It follows immediately that $\phi_{00}$ is stable if and only if
\[
D>\left.r_{1}^{g}\right|_{\phi_{00}},\;\left.r_{2}^{g}\right|_{\phi_{00}}.\]
 Since $\left.r_{1}^{g}\right|_{\phi_{00}}=r_{1}^{g}(\mathbf{s^{f}})$
and $\left.r_{2}^{g}\right|_{\phi_{00}}=r_{2}^{g}(\mathbf{s^{f}})$,
the stability condition says that the trivial steady state is asymptotically
stable if and only if the dilution rate exceeds the specific growth
rates of the two species at the feed concentrations.

\subsection{Semitrivial steady state}

The Jacobian at $\phi_{10}$ is \[
J(\phi_{10})=\left(\begin{array}{cccc}
-D-c_{1}\frac{\partial r_{11}^{s}}{\partial s_{1}} & -c_{1}\frac{\partial r_{11}^{s}}{\partial s_{2}} & -r_{11}^{s} & -r_{21}^{s}\\
-c_{1}\frac{\partial r_{12}^{s}}{\partial s_{1}} & -D-c_{1}\frac{\partial r_{12}^{s}}{\partial s_{2}} & -r_{12}^{s} & -r_{22}^{s}\\
c_{1}\frac{\partial r_{1}^{g}}{\partial s_{1}} & c_{1}\frac{\partial r_{1}^{g}}{\partial s_{2}} & 0 & 0\\
0 & 0 & 0 & r_{2}^{g}-D\end{array}\right).\]
 One of the eigenvalues of $J(\phi_{10})$ is \begin{equation}
\lambda_{2}=\left.r_{2}^{g}\right|_{\phi_{01}}-D.\label{eq:lambda1}\end{equation}
The sign of this eigenvalue tells us whether species 2 is capable
of invading the chemostat when it is operating at the steady state,
$\phi_{10}$. In particular, $\lambda_{1}<0$ if and only if $\left.r_{2}^{g}\right|_{\phi_{01}}<D$,
i.e., species 2 cannot invade the chemostat. However, this condition,
by itself, cannot ensure that $\phi_{10}$ is observable, since the
steady state might be unstable even if we deliberately exclude the
possibility of an invasion by species~2 by performing a single-species
experiment with species~1 at the same $D$ and $\mathbf{s_{f}}$.
The remaining three eigenvalues of $J(\phi_{10})$ tell us if $\phi_{10}$
is stable in such single-species experiments. Consistent with this
argument, the remaining three eigenvalues of $J(\phi_{10})$ are,
in fact, the eigenvalues of its submatrix \[
A=\left(\begin{array}{ccc}
-D-c_{1}\frac{\partial r_{11}^{s}}{\partial s_{1}} & -c_{1}\frac{\partial r_{11}^{s}}{\partial s_{2}} & -r_{11}^{s}\\
-c_{1}\frac{\partial r_{12}^{s}}{\partial s_{1}} & -D-c_{1}\frac{\partial r_{12}^{s}}{\partial s_{2}} & -r_{12}^{s}\\
c_{1}\frac{\partial r_{1}^{g}}{\partial s_{1}} & c_{1}\frac{\partial r_{1}^{g}}{\partial s_{2}} & 0\end{array}\right)\]
obtained by ignoring the existence of the second species. 

One of the eigenvalues of $A$ is $-D$, i.e.,\begin{equation}
\lambda_{2}=-D.\label{eq:lambda2}\end{equation}
 This is a consequence of the following {}``stoichiometric'' relationship
resulting from the assumption of constant yields\[
\frac{d}{dt}\left(Y_{11}s_{1}+Y_{12}s_{2}+c\right)=D\left(Y_{11}s_{1}^{f}+Y_{12}s_{2}^{f}\right)-D\left(Y_{11}s_{1}+Y_{12}s_{2}+c\right).\]
The signs of the remaining two eigenvalues of $A$, namely, can be
determined from the growth isocline and consumption curve for species
1. To see this, observe that \[
\mathnormal{\textrm{tr}}A=\lambda_{2}+\lambda_{3}+\lambda_{4}=-2D-c_{1}\frac{\partial r_{11}^{s}}{\partial s_{1}}-c_{1}\frac{\partial r_{12}^{s}}{\partial s_{2}}.\]
 It can also be shown that \[
\det A=\lambda_{2}\lambda_{3}\lambda_{4}=-c_{1}D\nabla r_{1}^{g}\cdot\mathbf{t}_{1}\]
 where $\mathbf{t}_{1}$ denotes the tangent at $\phi_{10}$ to the
consumption curve, $\Phi_{1}(\mathbf{s_{f}}),$ oriented along the
direction of increasing $s_{j}$. Since $\lambda_{2}=D$, we obtain
\begin{eqnarray}
\lambda_{3}+\lambda_{4} & = & -D-c_{1}\frac{\partial r_{11}^{s}}{\partial s_{1}}-c_{1}\frac{\partial r_{12}^{s}}{\partial s_{2}}<0,\label{eq:lambda34_1}\\
\lambda_{3}\lambda_{4} & = & c_{1}\nabla r_{1}^{g}\cdot\mathbf{t}_{1}\label{eq:lambda34_2}\end{eqnarray}
 so that the real parts of $\lambda_{3},\lambda_{4}$ are negative
if and only if $\nabla r_{1}^{g}\cdot\mathbf{t}_{1}>0$, i.e., the
angle between $\nabla r_{1}^{g}$ and \textbf{$\mathbf{t}_{1}$} is
less than $\pi/2$. Taken together, the relations, (\ref{eq:lambda1}--\ref{eq:lambda34_2})
imply that $\phi_{10}$ is stable if and only if $\left.r_{2}^{g}\right|_{\phi_{10}}<D$
and the angle between $\nabla r_{1}^{g}$ and $\mathbf{t}_{1}$ is
less than $\pi/2$.

A similar argument shows that $\phi_{01}$ is stable if and only if
$\left.r_{1}^{g}\right|_{\phi_{01}}<D$ and $\nabla r_{2}^{g}\cdot\mathbf{t}_{2}>0$,
where $\mathbf{t}_{2}$ is the tangent to the consumption curve at
$\phi_{01}$.

\subsection{Nontrivial steady state}

The Jacobian at $\phi_{11}$ is \[
J(\phi_{11})=\left(\begin{array}{cccc}
-D-c_{1}\frac{\partial r_{11}^{s}}{\partial s_{1}}-c_{2}\frac{\partial r_{21}^{s}}{\partial s_{1}} & -c_{1}\frac{\partial r_{11}^{s}}{\partial s_{2}}-c_{2}\frac{\partial r_{21}^{s}}{\partial s_{2}} & -r_{11}^{s} & -r_{21}^{s}\\
-c_{1}\frac{\partial r_{12}^{s}}{\partial s_{1}}-c_{2}\frac{\partial r_{21}^{s}}{\partial s_{2}} & -D-c_{1}\frac{\partial r_{11}^{s}}{\partial s_{1}}-c_{2}\frac{\partial r_{21}^{s}}{\partial s_{2}} & -r_{12}^{s} & -r_{22}^{s}\\
c_{1}\frac{\partial r_{1}^{g}}{\partial s_{1}} & c_{1}\frac{\partial r_{1}^{g}}{\partial s_{2}} & 0 & 0\\
c_{2}\frac{\partial r_{2}^{g}}{\partial s_{1}} & c_{2}\frac{\partial r_{2}^{g}}{\partial s_{2}} & 0 & 0\end{array}\right).\]
 It follows \[
\det\: J(\phi_{11})=c_{1}c_{2}\det(\mathbf{r_{1}^{s}},\mathbf{r_{2}^{s}})\cdot\det(\triangledown r_{1}^{g},\triangledown r_{2}^{g}).\]
 Since $c_{1},c_{2}>0$ at $\phi_{11}$, we conclude that a nontrivial
steady state is stable only if\begin{equation}
\det(\mathbf{r_{1}^{s}},\mathbf{r_{2}^{s}})\cdot\det(\triangledown r_{1}^{g},\triangledown r_{2}^{g})>0\label{necessary}\end{equation}
 i.e., $\det(\mathbf{r_{1}^{s}},\mathbf{r_{2}^{s}})$ and $\det(\triangledown r_{1}^{g},\triangledown r_{2}^{g})$
have the same sign, i.e., $(\mathbf{r_{1}^{s}},\mathbf{r_{2}^{s}})$
and $(\triangledown r_{1}^{g},\triangledown r_{2}^{g})$, have the
same orientations.

Note that the foregoing stability criterion is only a \emph{necessary}
condition for stability of a coexistence steady state~\citep{pilyugin03a}.
Hence, coexistence steady states at which $(\mathbf{r_{1}^{s}},\mathbf{r_{2}^{s}})$
and $(\triangledown r_{1}^{g},\triangledown r_{2}^{g})$, have opposite
orientations are certainly unstable. However, coexistence steady states
at which $(\mathbf{r_{1}^{s}},\mathbf{r_{2}^{s}})$ and $(\triangledown r_{1}^{g},\triangledown r_{2}^{g})$,
have the same orientation may or may not be stable.
\end{document}